\input{epsf}
\documentstyle[eqsecnum,aps,epsf]{revtex}

\begin{document}

\def\nbox#1#2{\vcenter{\hrule \hbox{\vrule height#2in
\kern#1in \vrule} \hrule}}  
\def\sq{\,\raise.5pt\hbox{$\nbox{.09}{.09}$}\,}
\def\sqb{\,\raise.5pt\hbox{$\overline{\nbox{.09}{.09}}$}\,}

\title
{Attractor states and infrared scaling in de Sitter space}
\author
{
Paul R. Anderson$^{1,2}$
\thanks{electronic address: anderson@wfu.edu},
Wayne Eaker$^{1}$
\thanks{electronic address: eaker@alumni.wfu.edu},
Salman Habib$^{2}$
\thanks{electronic address: habib@lanl.gov},
Carmen Molina-Par\'{\i}s$^{2,3}$
\thanks{electronic address: carmen@t6-serv.lanl.gov},
and
Emil Mottola$^{2}$
\thanks{electronic address: emil@lanl.gov}
}

\address{
$^{1}$
Department of Physics
Wake Forest University, Winston-Salem,
North Carolina, 27109\\
$^{2}$
T-8, Theoretical Division, Los Alamos
National Laboratory, Los Alamos, New Mexico, 87545
\\
$^{3}$
Centro de Astrobiolog\'\i a, CSIC-INTA, Carretera de Ajalvir Km. 4,
28850 Torrej\'o n, Madrid, Spain}

\preprint{LA-UR-00-52}
\date{LA-UR-00-52; May 23, 2000}
\maketitle
\begin{abstract}

{The renormalized expectation value of the energy-momentum tensor for
a scalar field with any mass $m$ and curvature coupling $\xi$ is
studied for an arbitrary homogeneous and isotropic physical
initial state in de Sitter spacetime. We prove quite generally that 
$\langle T_{ab}\rangle$ has a fixed point attractor behavior at late times, 
which depends only on $m$ and $\xi$, for any fourth order adiabatic state
that is infrared finite. Specifically, when $m^2+\xi R>0$, $\langle
T_{ab} \rangle$ approaches the Bunch-Davies de Sitter invariant value
at late times, independently of the initial state. When $m=\xi=0$, it
approaches instead the de Sitter invariant Allen-Folacci value. When
$m=0$ and $\xi\ge 0$ we show that this state independent asymptotic
value of the energy-momentum tensor is proportional to the conserved
geometrical tensor $^{(3)}H_{ab}$, which is related to the behavior of
the quantum effective action of the scalar field under global Weyl
rescaling.  This relationship serves to generalize the definition of
the trace anomaly in the infrared for massless, non-conformal
fields. In the case $m^2+\xi R=0$, but $m$ and $\xi$ separately
different from zero, $\langle T_{ab} \rangle$ grows linearly with
cosmic time at late times.  For most values of $m^2$ and $\xi$ in
the tachyonic cases, $m^2+\xi R < 0$, $\langle T_{ab} \rangle$ grows
exponentially at late cosmic times for all physically admissable
initial states. }

\end{abstract} 
\section{Introduction}

Quantum field theory in curved spacetimes does not contain in itself a
unique specification of the quantum state of the
system~\cite{bi-da}. Even in Minkowski spacetime, where the existence
of the Poincar\'e group singles out a special state, the Minkowski
vacuum, it is certainly of interest to consider states that are
non-invariant under Poincar\'e transformations, since they contain all
the information about the physical excitations and dynamics of the
theory. Such non-vacuum states are also necessary in a general initial
value formulation of the back-reaction problem in both curved and flat
spacetimes. In flat space the initial value problem for arbitrary
physically allowable states has been formulated and studied for both
QED and scalar $\Phi^4$ theory in the large $N$ limit, principally for
time varying but spatially homogeneous mean fields~\cite{chkm,HKMP}.

The simplest situation in which the back-reaction problem can be
studied in curved spacetime is that of a free scalar field in a
spatially homogeneous and isotropic Robertson-Walker (RW) cosmology,
where the geometry is characterized by just one non-trivial function
of time. The wave equation for a free scalar field in such a geometry
can be separated and expressed in terms of a complete set of time
dependent mode functions. The general initial value problem is
specified by giving initial data for this complete set at a given
initial time. The back-reaction of the quantum scalar field(s) on the
RW geometry can be studied then by constructing the renormalized
expectation value of the energy-momentum tensor $\langle T_{ab}
\rangle$ of the field(s) and solving (numerically) the semi-classical
Einstein equations, augmented by higher derivative terms required by
renormalization\cite{wald,star,fhh,a1,wada,sa}.  As in the flat space
examples, this semi-classical back-reaction problem becomes exact in
the large $N$ limit, with $N$ the number of identical scalar
fields~\cite{tom}.

As a prelude to the dynamical back-reaction problem in cosmological
spacetimes it is necessary to study non-vacuum states first in fixed
RW backgrounds. The maximally symmetric de Sitter spacetime is of
particular interest. Most previous work has focused on maximally
symmetric $O(4,1)$ de Sitter invariant states or the special $O(4)$
invariant state found by Allen ~\cite{a}.  Since the universe is not
globally $O(4,1)$ invariant, a more generic set of initial conditions,
consistent only with RW symmetry and general principles of
renormalization of $\langle T_{ab} \rangle$ is required for
cosmology. The investigation of these much weaker requirements and
specification of the general initial value problem for back-reaction
calculations was initiated in Ref.~\cite{hmp}.

In this paper we study the behavior of the renormalized $\langle
T_{ab} \rangle$ for arbitrary physically admissable spatially
homogeneous and isotropic states in a fixed de Sitter background.  We
argue in Section III that such states must be fourth order adiabatic
states~\cite{bi-da} that also possess an infrared finite two-point function. 
In de Sitter space the wave equation for free scalar fields can be solved 
exactly for arbitrary values of the mass and the curvature coupling. Its solutions
depend only on the wave number $k$ of the mode and the parameter
$\nu^2 ={9\over 4}-m^2 \alpha^2 -12\xi$, with $R=12\alpha^{-2}$ the constant
scalar curvature of de Sitter spacetime.  For $\Re (\nu)<{3\over 2}$,
corresponding to $m^2+\xi R>0$, we prove that for {\it all} UV and IR 
physically allowed initial states the renormalized value of $\langle T_{ab}
\rangle$ at late times asymptotically approaches that of the Euclidean
or Bunch-Davies de Sitter invariant state~\cite{c-t,t,d-c,bu-da}.  The
conformally invariant scalar field ($m=0$,\ $\xi={1\over 6}$) falls
into this class.

The case $\nu = {3\over 2}$ corresponding to $m^2 \alpha^2 +12\xi =0$
is more delicate. If $m$ and $\xi$ are separately zero (the massless,
minimally coupled case), then we prove that the renormalized $\langle
T_{ab} \rangle$ for all physically admissable states approaches the
Allen-Folacci de Sitter invariant value~\cite{a-f,f,k-g}. Numerical
evidence for this result was found previously in Ref.~\cite{hmp}.  In
this paper we provide an analytic proof that late time attractor behavior
occurs for {\it all} physically admissable RW states, when $m^2 \ge 0$ and
$\xi \ge 0$.  If $m^2 \alpha^2 +12\xi =0$ but $m^2$ and $\xi$ are not
separately zero (so that one of them is negative) we prove that
$\langle T_{ab} \rangle$ grows linearly in RW comoving (cosmic) time
without bound, and this asymptotic behavior is independent of the
state of the field. Finally, and in contrast, in the case $\nu>{3\over
2}$, corresponding to $m^2+\xi R<0$, $\langle T_{ab} \rangle$ depends
sensitively on the state and, for most values of $m$ and $\xi$, grows
exponentially at late times for all states.  This case is of
considerably less physical relevance, since it corresponds to a
tachyonic field theory with no stable vacuum state.

The asymptotic approach of $\langle T_{ab} \rangle$ to a de Sitter
invariant form, independently of the lower symmetry of the initial
data when $\Re(\nu)\le {3\over 2}$ is a striking result. Certainly no
such attractor behavior of $\langle T_{ab} \rangle$, independent of
initial conditions occurs in Minkowski space for any mass. One may
regard this result as a kind of cosmic ``no hair'' theorem for scalar
quantum fields in de Sitter space.  For $\Re(\nu) < {3\over 2}$ it is
in accord with one's classical intuition that any initial energy
density satisfying the weak energy condition ($\varepsilon +p>0$) is
redshifted away by the exponential de Sitter expansion, although as we
will see, the redshifting of the quantum $\langle T_{ab} \rangle$ is
not that of classical matter or radiation.  At asymptotically late
times what is left behind is a kind of frozen quantum vacuum energy
``condensate,'' satisfying the de Sitter invariant equation of state
$p=-\varepsilon$. This result justifies the choice of the Bunch-Davies
vacuum in calculations of quantum fluctuations of {\it free} fields,
{\it i.e.} without back-reaction, in a long-lived de Sitter expansion
phase of inflationary cosmological models. For $\nu = {3\over 2}$, $m
= \xi = 0$, the approach of $\langle T_{ab} \rangle$ to the de Sitter
invariant Allen-Folacci value is perhaps more surprising.  As shown in
Section IV one expects the leading order contribution of the modes to
$\langle T_{ab} \rangle$ in this case to be constant in comoving time
at late times.  In fact this occurs if $m$ and $\xi$ are not
separately zero for all the modes.  However when $m$ and $\xi$ are
both zero, the leading order contributions to $\langle T_{ab} \rangle$
of all the modes {\it except the spatially homogeneous one}, for an
arbitrary physically admissable state have exactly zero coefficient, the
subleading contributions redshift away, and we are left only with the
with the de Sitter invariant Allen-Folacci constant value at late
times. The finite difference from the Bunch-Davies value may be
attributed entirely to the constant behavior of the spatially
homogeneous mode contributing to the vacuum energy condensate in de
Sitter space.

In all those cases for which $\Re(\nu)\le {3\over 2}$ when $\langle
T_{ab} \rangle$ approaches a de Sitter invariant value at late times,
the quantum expectation value loses all its initial state dependence
and hence its asymptotic value must be determined purely by the
background geometry. When the mass of the field vanishes, the
existence of only one covariantly conserved local geometrical tensor
of adiabatic order four in de Sitter space, namely $^{(3)} H_{ab}$
given by Eq. (\ref{eq:Hthr}) below, permits us to identify the
asymptotic value of $\langle T_{ab} \rangle$ in the vacuum energy
condensate with this tensor. Since $^{(3)}H_{ab}$ cannot be derived by
variation of a covariant local action, but corresponds instead to a
certain well defined non-local term in the quantum effective
action~\cite{amm}, the asymptotic vacuum energy condensate of the
quantum field is determined by the global or extreme infrared
properties of de Sitter space. The form of the non-local effective
action is determined by the trace anomaly in conformally flat
spacetimes. Since the approach of $\langle T_{ab} \rangle$ to a de
Sitter invariant value occurs for all $\xi \ge 0$, the existence of
this term in the effective action for massless fields is much more
general than the strict definition of the trace anomaly in the
conformally invariant case. Hence the asymptotic late time behavior of
$\langle T_{ab} \rangle$ in de Sitter space can be used to define a
generalized trace ``anomaly'' coefficient in the massless but
non-conformally invariant cases, $\xi \neq {1\over 6}$. As we show by
consideration of the covariant $\zeta$ function method~\cite{bi-da},
this coefficient is exactly the same as that which determines the
infrared response of the vacuum condensate to global Weyl
rescalings. Hence the significance of the state-independence of the
vacuum energy condensate in de Sitter space is that it determines
certain conformal properties of non-conformal field theories in the
extreme infrared.

The paper is organized as follows. In Section \ref{sec:scalar} we give
the expectation value of the energy-momentum tensor as a mode sum for
an arbitrary homogeneous and isotropic, physically admissable state 
with a non-zero initial particle number of the scalar field. In Section 
\ref{sec:analytic} we analyze the late time behavior of expectation value 
of the energy-momentum tensor in de Sitter space in flat spatial sections 
and show that it approaches the de Sitter invariant Bunch-Davies value for
all $\Re (\nu) < {3\over 2}$, independently of the initial state. In
Section \ref{sec:nu32} we analyze the limit $\Re (\nu) \rightarrow
{3\over 2}$ in closed spatial sections in order to keep careful track
of the spatially homogeneous mode in a discrete basis, and show how
the Allen-Folacci fixed point at late times is obtained for the
massless minimally coupled field. We also investigate the asymptotic
behavior of the energy-momentum tensor for arbitrary mass and
curvature coupling when $\nu \ge {3\over 2}$.  In Section \ref{sec:numerical}
we illustrate the analytic results with numerical studies,
investigating in particular the interesting case when $\nu$ is
slightly smaller than ${3\over 2}$. We find that for many states when
$\nu$ is only slightly smaller than ${3\over 2}$ the energy-momentum
tensor first approaches the Allen-Folacci value and only much later
approaches the Bunch-Davies value. In Section \ref{sec:scaling} we
consider the geometric significance of the state independent
asymptotic behavior of $\langle T_{ab} \rangle$, relating it to the
quantum effective action which determines the behavior of $S_{\mathrm
eff}$ under global Weyl scaling and providing the generalization of
the trace ``anomaly'' in the infrared, for $\xi \neq {1\over
6}$. Section \ref{sec:discussion} contains some discussion and final
conclusions. There are two Appendices. Appendix A completes the proof
of the Bunch-Davies attractor behavior in the cases of integer and
pure imaginary $\nu$, while Appendix B contains a discussion of the
simple harmonic oscillator in the limit of vanishing frequency, which
shares many features with the spatially constant mode in de Sitter
spacetime.

\section{Scalar field in a RW background}
\label{sec:scalar}

The metric for a general RW spacetime can be written in conformal time
$\eta$ in the form
\begin{equation}
{\mathrm{d}}s^2 = a^2(\eta) \left(-{\mathrm{d}}\eta^2
+ \frac{{\mathrm{d}}r^2}{1 - \kappa r^2}
           + r^2 {\mathrm{d}} \Omega^2 \right)
\; . \label{eq:metricRW}
\end{equation}
Here $a(\eta)$ is the scale factor and $\kappa =0,+1,-1$
corresponds to the cases of flat, spherical, and hyperbolic spatial
sections, respectively. Throughout we use units such that
$\hbar = c = 1$ and the Misner, Thorne, and Wheeler \cite{MTW} conventions 
for the curvature tensors,
$R^a_{bcd} = \Gamma^a_{bd,c} - ...$ and $ R_{ab} = R^c_{acb}$.

We consider in this paper a free quantum scalar field $\Phi$ with the
quadratic action
\begin{eqnarray}
S =  -{1\over 2}\int\,
{\mathrm{d}}^4 x \sqrt{-g}
\, \left[(\nabla_a \Phi)g^{ab}
(\nabla_b \Phi)
+ m^2\Phi^2  + \xi R \Phi^2\right]
\; ,
\label{action}
\end{eqnarray}
where $\nabla_a$ denotes the covariant
derivative, $R$ is the scalar curvature, and $g\equiv \det(g_{ab})$. The
mass $m$ and curvature coupling $\xi$ are allowed to have any real value. 
The wave equation for $\Phi$ obtained by varying this action is
\begin{eqnarray}
\left[-{\,\raise.5pt\hbox{$\mbox{.09}{.09}$}\,}
+ m^2 + \xi R\right] \Phi(\eta, {\bf x}) = \left[ {1\over a^4} {\partial\over
\partial \eta}\left(a^2 {\partial\over \partial \eta}\right) - \frac{1}{a^2}
 \Delta^{(3)} + m^2 + \xi R\right]\Phi = 0\; ,
\label{waveq}
\end{eqnarray}
with $\Delta^{(3)}$ the covariant spatial Laplacian.
For spacetimes with the metric (\ref{eq:metricRW}) the field $\Phi$ can be
expanded as a mode sum in the form,~\cite{bi-da}
\begin{equation}
\Phi(\eta,{\bf x}) = \frac{1}{a(\eta)} \int {\mathrm{d}}
\tilde\mu({\bf k}) \left[ a_{\bf k} Y_{\bf k}({\bf x}) \psi_k (\eta)
          + a_{\bf k}^\dagger Y_{\bf k}^*({\bf x})
\psi^*_k (\eta) \right]
\; ,
\end{equation}
where the integration measure is given by
\begin{eqnarray}
  \int {\mathrm{d}} \tilde\mu({\bf k}) 
\equiv
\left\{
\begin{array}{lll}
\int {\mathrm{d}}^3 {\bf k} \;\; & {\rm if}& \kappa = 0 \; ,
\vspace{0.2cm}
 \nonumber \\
  \int_0^{\infty} {\mathrm{d}} k \sum_{l,m} \;\; & {\rm if}&
\kappa = -1 \; ,
\vspace{0.2cm}
\nonumber \\
 \sum_{k,l,m} \;\; & {\rm if} & \kappa = +1 \; ,
\end{array}
\right.
\end{eqnarray}
and the spatial part of the mode functions $Y_{\bf k}({\bf x})$ obeys the
equation
\begin{equation}
-\Delta^{(3)} Y_{\bf k}({\bf x}) = (k^2 - \kappa) Y_{\bf k}({\bf x})
\; ,
\end{equation}
with $k = 1, 2, \dots$ in the case of closed spatial sections, $\kappa = +1$.
The time dependent part of the mode functions $\psi_k$ obeys the equation
\begin{equation}
{\psi_k}'' + \left[k^2 + m^2 a^2 + \left(\xi 
- {1\over 6}\right) a^2 R\right] \psi_k = 0
\;,
\label{eq:mode}
\end{equation}
where primes denote derivatives with respect to the conformal time
variable $\eta$, and the scalar curvature in a general RW spacetime is
given by
\begin{equation}
R = 6 \left(\frac{a''}{a^3} + \frac{\kappa}{a^2} \right)
\;.
\end{equation}
For the quantum field to satisfy the canonical commutation
relations, the creation and annihilation operators are required to obey
the commutation relations $[a_{\bf k},a^{\dagger}_{\bf k'}]=
\delta_{{\bf k}{\bf k'}}$, whereupon the $\psi_k$ must obey the Wronskian 
condition
\begin{equation}
 \psi_k {\psi^{*}_k}' - \psi^*_k \psi'_k = i\,.   \label{eq:wronskian}
\end{equation}

The components of the unrenormalized energy-momentum tensor (energy density
and trace) are given by~\cite{Bunch}
\begin{mathletters}
\begin{eqnarray}
\varepsilon_{u} &=& -\langle {T^0}_0\rangle_u =
\frac{1}{4 \pi^2 a^4}\int {\rm{d}}\mu(k)
(2 n_k + 1) \left\{|\psi'_k|^2 + (k^2+m^2 a^2) |\psi_k|^2\right.
\nonumber \\
& &\left.\;\;\;\; + \; \left(6\xi - 1\right) \left[\frac{a'}{a} 
(\psi_k {\psi^{*}_k}' + \psi_k^* \psi'_k) - \left(
\frac{a^{\prime 2}}{a^2}- \kappa\right)
|\psi_k|^2\right]\right\} \; ,
 \label{eq:T00u}
\end{eqnarray}
and
\begin{eqnarray}
-\varepsilon_{u} + 3p_{u} &=& \langle T\rangle_{u} =
\frac{1}{2 \pi^2 a^4}\int {\rm{d}}
\mu(k)
 (2 n_{k} + 1)
\left\{ -m^2 a^2 |\psi_k|^2
     + \left(6\xi - 1\right)\left[-|\psi'_k|^2
      + \frac{a'}{a} (\psi_k  {\psi^{*}_k}'  + \psi_k^* {\psi'_k})
\right]\right.
 \nonumber \\
  &  &
     \left.+ \left(6\xi-1\right)\left[k^2+m^2 a^2 + \left(\frac{a''}{a}-
\frac{a'^2}{a^2}\right) + \left( \xi-\frac{1}{6} \right) a^2
R\right] |\psi_k|^2 ]\right\}  \label{eq:Tru}
\; .
\end{eqnarray}
\end{mathletters}%
where we have allowed an arbitrary number of particles in the initial
state, $n_k = \langle a^{\dagger}_{\bf k} a_{\bf k}\rangle$, and the
scalar measure ${\mathrm{d}}\mu (k)$ is given by
\begin{eqnarray*}
 \int {\mathrm{d}} \mu(k) \equiv
\left\{
\begin{array}{lll}
 \int_0^{\infty}
 {\mathrm{d}} k \;  k^2 \;\;\;
& {\rm if}& \kappa = 0, -1 \; ,
\vspace{0.2cm}
\nonumber \\
       \sum_{1}^{\infty} k^2 \;\;\; &{\rm if}&
\kappa = +1 \; .
\end{array}
\right.
\end{eqnarray*}
As we are considering spatially homogeneous and isotropic initial
states (consistent with the RW symmetry), $n_k$ depends only on the
magnitude $k$ of the spatial wave vector $\bf k$. Expectation values
of the bilinears $\langle a_{\bf k} a_{\bf k}\rangle$ and $\langle
a^{\dagger}_{\bf k} a^{\dagger}_{\bf k}\rangle$ in a general state
need not be considered since they can be removed by a time-independent
Bogoliubov transformation at the initial time~\cite{chkm}.  Hence
these initial state correlations may be parameterized instead by the
initial data on the mode functions $\psi_k$, together with the
non-negative set of $n_k$, with no loss of generality.

Since the expectation value of the unrenormalized energy-momentum
tensor $\langle T_{ab}\rangle_{u}$ is quartically divergent, a
procedure for defining a finite, renormalized expectation value must
be given. We will follow the adiabatic regularization
method~\cite{P,ParkerFulling,FP,FPH}. In this method the
renormalization counterterms are constructed using a fourth order WKB
expansion for the mode functions.  We denote these counterterms by 
$\langle T_{ab}\rangle_{ad}$. They are given in Refs.~\cite{Bunch} and~\cite{AP}.  
The renormalized energy-momentum tensor is then
\begin{equation}
\langle T_{ab}\rangle_{ren} \,=\, \langle T_{ab}\rangle_{u} -
\langle T_{ab}\rangle_{ad}
\;.
\label{eq:trenorm}
\end{equation}
This subtraction scheme is not manifestly covariant in form, since space
and time are treated quite differently.  However, adiabatic
regularization is equivalent to a covariant point splitting procedure
in which the points are split only in the spacelike hypersurface of
constant $\eta$~\cite{AP,Birr}, and the {\it value} of the
renormalized $\langle T_{ab} \rangle$ obtained by this procedure is
the same as in a strictly covariant one. Hence this subtraction
procedure does correspond to adjustment of counterterms to
the quantum effective action, and $\langle T_{ab}\rangle_{ren}$ is
covariantly conserved. As discussed in detail in
Ref.~\cite{AP} the adiabatic terms in all cases consist of
an integral rather than a sum over $k$. The reason is that
subtraction corresponds to purely local counterterms in the effective
action, and thus must be independent of the global compactness or
non-compactness of the spatial sections.

For an arbitrary homogeneous and isotropic state to be physically admissible
the renormalized energy-momentum tensor defined by the adiabatic order
four subtractions in (\ref{eq:trenorm}) must be both ultraviolet and
infrared finite. In a general RW spacetime, ultraviolet finiteness for 
a field with non-conformal
coupling to the scalar curvature requires that the particular
solution of the mode equation (\ref{eq:mode}) in a general physical
state must match the fourth order adiabatic form at large $k$,
with the deviations from the fourth order WKB form falling faster than
$k^{-4}$. Likewise the initial number $n_k$ must fall faster than
$k^{-4}$ at large $k$, for the mode sums (integrals) to be ultraviolet
convergent.  This is equivalent to the requirement that the two-point
function of the scalar field have the vacuum Hadamard 
form~\cite{r-l,n-o,junker,lindig}
to sufficiently high order in the short distance expansion as the points
approach one another.  As long as there are two linearly independent
complex oscillatory solutions to the equation (\ref{eq:mode}), the Wronskian
normalization condition (\ref{eq:wronskian}) can be imposed and the
state will be free of any infrared divergences. However, when
$m^2 + \xi R \rightarrow 0$ for some low $k$ in de Sitter space 
no such complex oscillatory solutions to (\ref{eq:mode}) exist 
and the infrared finiteness requirement on the both the energy-momentum 
tensor and the two-point function of the physical state becomes 
non-trivial, as we discuss in detail in Section IV.

A useful variation of the method of adiabatic regularization has
been developed by two of us~\cite{AE}.  In this method
one first computes a quantity $\langle {T_{ab}}\rangle_{d}$,
obtained by expanding the adiabatic counterterms $\langle
{T_{ab}}\rangle_{ad}$ in inverse powers of $k$ and truncating
at order $k^{-3}$. The same renormalized energy-momentum tensor
defined in Eq. (\ref{eq:trenorm}) is separated into the sum
of two {\it finite} terms by adding and subtracting the simplified
form of the divergent counterterms $\langle {T_{ab}}\rangle_{d}$
\begin{eqnarray}
 \langle {T_{a b}}\rangle_{ren} &=& \langle {T_{ab}}\rangle_{n}
+ \langle {T_{ab}}\rangle_{an}
\; ,    \nonumber \\
 \langle {T_{ab}}\rangle_{n} &=& \langle {T_{ab}}\rangle_{u}
- \langle {T_{ab}}\rangle_{d}
\; , \nonumber \\
 \langle {T_{ab}}\rangle_{an} &=& \langle {T_{ab}}\rangle_{d}
- \langle {T_{ab}}\rangle_{ad}
\; .             \label{eq:deftan}
\end{eqnarray}
The full expressions for $\langle T_{ab}\rangle_d$ and $\langle
T_{ab}\rangle_{an}$ are given in Ref.~\cite{AE} for a general
RW spacetime. The advantage of this splitting
is that $\langle T_{ab}\rangle_{n}$ and $\langle T_{ab}\rangle_{an}$
are separately conserved, and moreover, $\langle T_{ab}\rangle_{an}$
may be computed analytically in terms of the scale factor
$a(\eta)$ and its derivatives~\cite{AE}. Thus the state dependence of
the renormalized $\langle T_{ab}\rangle_{ren}$ resides completely in
$\langle T_{ab}\rangle_{n}$, which can be computed numerically.

\section{The Bunch-Davies Attractor for $\Re (\nu) <{3\over 2}$}
\label{sec:analytic}

We restrict our consideration henceforth to the particular maximally
symmetric RW background of de Sitter space. The geometry of de Sitter
spacetime can be described in a number of different coordinate
systems. If $\kappa =0$ the spatial sections are flat and the
scale factor is
\begin{equation}
a(\eta) =- \frac{\alpha}{\eta} \;,\;\;\; -\infty < \eta < 0\; ,
\;\;\; \kappa =0 \; ,
\label{eq:ak0}
\end{equation}
with $\alpha$ a real, positive constant, and $R = 12\alpha^{-2}$.
If $\kappa =+1$ then the scale factor is
\begin{equation}
a(\eta) = \alpha\,\sec\eta \; ,\;\;\;\;
-\frac{\pi}{2} < \eta < \frac{\pi}{2} \;, \;\;\;  \kappa =+ 1 \; ,
\label{eq:ak1}
\end{equation}
which is equivalent to $a(\eta)=\alpha \csc \eta$ with $0 < \eta < \pi$.  To
simplify the notation we will generally use dimensionless units where
$\alpha = 1$ and $R= 12$, restoring the dimensionful quantities
when it is instructive to do so.

The asymptotic behavior of the energy-momentum tensor does not depend
on $\kappa$ so the bulk of the analysis will be carried out in the
flat ($\kappa = 0$) coordinates. However, when we turn to the massless,
minimally coupled limit in the next section, it will become useful to
have a discrete $k$ basis in order to separate out the $k=1$ spatially
homogeneous mode explicitly, since it is the most infrared sensitive.  
No confusion should be caused by our use of the same symbol $\eta$ for 
conformal time in both cases of flat and closed spatial sections, 
since we make use of only the $\kappa = 0$ coordinates in this Section 
and only the $\kappa = +1$ coordinates in the next Section. 
We will not make use of the spatially open ($\kappa = -1$) coordinates 
in this paper.

For the case of Eq. (\ref{eq:ak0}) the general solution to the mode
equation can be written as~\cite{bu-da} \footnote{In Ref.~\cite{bu-da}
the arguments of the Hankel functions are given as $k\eta$ rather than
$-k\eta$. We have chosen to use non-negative arguments to avoid
complications that result from the fact that these functions have
branch cuts along the negative real axis.}
\begin{equation}
\psi_k(\eta) =
\frac{1}{2} (-\pi\eta)^{1\over 2} e^{\frac{i\nu\pi}{2}} 
\left[c_1(k) H_\nu^{(1)} (-k\eta)
       + c_2(k) H_\nu^{(2)}(-k\eta) \right]
\; ,
    \label{eq:psiK0}
\end{equation}
where the $H_\nu^{(1),(2)}$ are Hankel functions and
\begin{equation}
\nu^2 \equiv \frac{9}{4} - m^2 \alpha^2 - 12 \xi \equiv -\gamma^2
\; .
 \label{eq:nu}
\end{equation}
The latter notation is useful in the case $\nu^2 <0$ so that $\nu =
i\gamma$ is purely imaginary. When $\nu^2 > 0$ we will choose $\nu$ to
be the positive root of (\ref{eq:nu}). From Eq.\ (\ref{eq:psiK0}) we
see that solutions to the mode equation in de Sitter space depend on
$m$ and $\xi$ only through their dependence on the parameter
$\nu$. Note that because of the minus sign in the arguments of the
Hankel functions, it is the function $H_\nu^{(1)}$ that corresponds to
a positive frequency mode in the large $k$ limit. The normalization of
the mode function in (\ref{eq:psiK0}) has been chosen so that the Wronskian
condition (\ref{eq:wronskian}) becomes simply
\begin{equation}
|c_1(k)|^2 - |c_2(k)|^2 = 1
\; .
\label{eq:c1c2K0}
\end{equation}
The Bunch-Davies (BD) state is defined by the choice, $c_1 = 1$ and $c_2 = 0$ 
(with $n_k =0$) for all $k$. The renormalized energy-momentum tensor in 
the BD state is given by~\cite{d-c,bu-da}
\begin{eqnarray}
\langle T_{ab}\rangle_{BD} &=&
-\frac{g_{ab}}{64 \pi^2} \left\{m^2\left[m^2 +
 \left(\xi - \frac{1}{6}\right) R\right]
\left[\psi \left(\frac{3}{2} + \nu \right)
+
\psi \left(\frac{3}{2} - \nu \right)
- \log \left(\frac{12 m^2}{R}\right) \right]
\right.
\nonumber \\
&-& \left.
m^2 \left(\xi-\frac{1}{6}\right) R
 - \frac{1}{18} m^2 R
- \frac{1}{2} \left(\xi - \frac{1}{6}\right)^2 R^2
+ \frac{R^2}{2160} \right\}
\; ,
\label{eq:tmunu-bd}
\end{eqnarray}
where $\psi(z) = {d \log \Gamma (z)\over dz}$ is the digamma function.

That this finite value of $\langle T_{ab}\rangle_{BD}$ coincides with
the renormalized $\langle T_{ab}\rangle_{ren}$ defined by the
adiabatic subtraction in (\ref{eq:trenorm}) follows from the fact that
the BD state is an allowed fourth order adiabatic state.  This may be
checked by comparing the asymptotic expansion of the exact BD mode
function, $\frac{(-\pi\eta)^{1\over 2}}{2}H_\nu^{(1)}(-k\eta)$, for
large $k$ with the fourth order adiabatic mode function
\begin{equation}
\psi_k^{(4)}(\eta) = \frac {1} {\sqrt {2 W_k^{(4)}}} 
\exp\left( - i\int^{\eta}\,
W_k^{(4)}(\eta')\,{\rm d}\eta'\right)\,,
\label{eq:psiadb}
\end{equation}
with $W_k^{(4)}(\eta)$ the fourth order adiabatic frequency.  It is given
explicitly in de Sitter space in flat conformal time coordinates by
\begin{equation}
W_k^{(4)}(\eta) = k + {1 \over 2 k\eta^2}\left(\frac{1}{4} - \nu^2\right)
- {1 \over 8 k^3 \eta^4}\left(\frac{1}{4} - \nu^2\right)
\left(\frac{25}{4} - \nu^2\right) + {\cal O}\left( {1\over k^5}\right)\,,
\end{equation}
up to the required order at large $k$. 

For the general state with $c_2 \neq 0$ to remain fourth order
adiabatic, we must have for large values of $k$
\begin{equation}
 c_2(k) = \frac{C(k)}{k^4} \; , \label{eq:ckbound} \end{equation} for
some complex function $C(k)$ which vanishes in the limit $k\rightarrow
\infty$. This condition is necessary for an arbitrary (spatially
homogeneous) state to posses a finite energy-momentum tensor after the
fourth order adiabatic subtraction defined by
(\ref{eq:trenorm}). Likewise the same condition of finite $\langle
T_{ab}\rangle$ requires us to restrict the average number of particles
$\langle a^{\dagger}_{\bf k} a_{\bf k}\rangle = n_k$ by
\begin{equation} n_k = \frac{N(k)}{k^{4}} \; , \label{eq:nkbound}
\end{equation} for some real function $N(k)$ which vanishes in the
limit $k\rightarrow \infty$.  The two ultraviolet conditions
\begin{equation} \lim_{k\rightarrow \infty} \vert C(k)\vert =
\lim_{k\rightarrow\infty} N(k) =0\,, \label{eq:CNlim} \end{equation}
on the physically allowed states guarantee that the Green's function
for the scalar field is locally of the Hadamard
form~\cite{r-l,n-o,junker,lindig}, and that the divergences of
$\langle T_{ab} \rangle$ match those of the fourth order adiabatic
vacuum, and are removed by the adiabatic subtraction procedure. We
will call any state which satisfies the conditions (\ref{eq:ckbound}),
(\ref{eq:nkbound}), and (\ref{eq:CNlim}), together with the Wronskian
condition (\ref{eq:c1c2K0}), a UV admissible physical
state~\footnote{In fact, the requirement of fourth order adiabatic
states is slightly more restrictive than UV finiteness of the
energy-momentum tensor in de Sitter space, since $C(k)$ can go to a
non-zero constant at large $k$ and the mode sums still converge. This
is associated with the vanishing of the tensor, $^{(1)}H_{ab}$,
defined in Eq. (\ref{eq:Hone}) in de Sitter space. If the field is
non-conformally coupled, a state with $C(k) \rightarrow $ constant would
lead to a divergent energy-momentum tensor if the spacetime is not
exactly de Sitter. If the field is conformally coupled and the
spacetime is not exactly homogeneous and isotropic, then the
energy-momentum tensor would again be divergent. Thus the most general
physically acceptable UV states are fourth order adiabatic states.}.

We shall require also that the arbitrary physical state possess a
two-point function and energy-momentum tensor which are free of any
infrared divergences.  Because the canonical dimension of $\langle
T_{ab}\rangle$ is four whereas that of $\langle
\Phi(x)\Phi(x')\rangle$ is two, the conditions (\ref{eq:ckbound}) and
(\ref{eq:nkbound}) which require finiteness of $\langle T_{ab}
\rangle_{ren}$ are more restrictive in the UV, whereas the condition
of finiteness of $\langle \Phi(x)\Phi(x')\rangle$ is more restrictive
in the IR.  These two sets of conditions will be sufficient to
demonstrate that the energy-momentum tensor for {\it any} UV and IR
admissible physical state approaches the BD value at late times for
$\Re (\nu)< {3\over 2}$.

To understand why such a result is to be expected and outline the
more detailed proof which we give below, let us observe that at late times 
$\eta \rightarrow 0^-$, the general state mode function 
(\ref{eq:psiK0}) behaves like
\begin{equation}
\psi_k \sim (-\eta)^{{1\over 2}\, -\nu} \sim a^{\nu -{1\over 2}} \; .
\label{eq:psilate}
\end{equation}
Substituting this into (\ref{eq:T00u}) and (\ref{eq:Tru}) shows that
to leading order at late times the contributions to the mode sums of
the unrenormalized energy-momentum tensor behave like
$(-\eta)^{3-2\nu} \sim a^{2\nu -3}$ for $\nu$ real.  Since the
renormalization counterterms are state independent~\cite{Bunch}, the
state dependent terms are the same in the unrenormalized and
renormalized energy-momentum tensor.  One can perform all the UV
renormalization in the BD state at a fixed time and collect the
remaining finite state dependent terms which are unaffected by the
subtraction procedure, and they all fall off at least as fast as
$(-\eta)^{3-2\nu}$ as $\eta \rightarrow 0^-$ for $\Re (\nu) < {3\over
2}$.

These remaining finite state dependent terms in the energy density and
trace are expressible as integrals over the wave number $k$ with the
general form
\begin{equation}
I (\eta) = \int_0^{\infty}\, {{\rm d}k \over k}\ R(k)\ S(-k\eta)\,,
\label{eq:genint}
\end{equation}
where $R(k)$ is one of the four state dependent, but time independent functions
\begin{eqnarray}
 \vert c_2(k)\vert^2 &&(1 + 2n_k),\nonumber\\
 \Re [c_1(k) c_2(k)]&&(1 + 2n_k),\nonumber\\
 \Im [c_1(k) c_2(k)]&&(1 + 2n_k), \nonumber\\ && n_k, \end{eqnarray}
and $S(z=-k\eta)$ is a product of the state independent Bessel
functions and their derivatives. An explicit basis for the twelve
products of Bessel functions $S_i (z)$ for $i = 1,\dots, 12$ which
appear in the integrals is given in Table 1. The essential point
is that all the state dependent mode integrals of the form
(\ref{eq:genint}) are {\it uniformly} convergent for all $\eta$
(including $\eta =0$) at both their lower limit, $k=0$, and their upper
limit $k=\infty$, due to the IR and UV finiteness of the state. Hence
the limit of $\eta \rightarrow 0^-$ can be taken {\it inside} the
integral over $k$. Since, as Table 1 shows, all the $S_i(z)$ behave
like 
\begin{equation} 
S_i(z) \rightarrow s_{i, 0}\ z^{\beta_i}\,,
\qquad {\rm with} \qquad \beta_i > 0\,, 
\label{eq:Slim} 
\end{equation}
as $z\rightarrow 0$, for $\Re (\nu) < {3\over 2}$, it follows that
\begin{equation} 
\lim_{\eta\rightarrow 0^{-}} I(\eta) =
\int_0^{\infty}\, {{\rm d}k \over k}\ R(k)\ \lim_{\eta\rightarrow
0^{-}} S(-k\eta) = 0\, , 
\end{equation} and all the state dependent
contributions to $\langle T_{ab} \rangle_{ren}$ vanish at late times.

The validity of bringing the limit inside the integral depends on the
uniform convergence of the integral at both its upper and lower
limits. In the form (\ref{eq:genint}) the behavior of the $S(-k \eta)$
factor at small arguments (and the absence of any IR divergence from
the $R(k)$ factor) clearly guarantees the uniform convergence at the
lower limit. However, the change of variables $z= -k\eta$ brings the
integral (\ref{eq:genint}) into the form 
\begin{equation} 
I(\eta) =
\int_0^{\infty}\, {{\rm d}z \over z}\ R\left({z\over -\eta}\right)\ S(z)\,.
\label{eq:genintz} 
\end{equation} 
In this form it is clear that the
uniform convergence of the integral at its upper limit is guaranteed
by the falloff of the state dependent mode functions at large $k$,
namely
\begin{equation} \lim_{k \rightarrow \infty} R(k) =
\lim_{\eta \rightarrow 0^-}R \left({z\over -\eta}\right) = 0
\; .
\label{eq:Rlim} 
\end{equation} 
Equations (\ref{eq:Slim}) and
(\ref{eq:Rlim}) guarantee that both the IR and UV contributions go to
zero as $\eta\rightarrow 0^{-}$ for $\Re (\nu) < {3\over 2}$.

In fact, we can go one step further by using the information from the
fourth order adiabatic nature of the state 
\begin{equation} \lim_{k
\rightarrow \infty} k^4 R(k) = 0\,, 
\label{eq:adbR} 
\end{equation} to
conclude that the UV contribution to the integral from $z \sim 1$, and
very large $k \sim (-\eta)^{-1}$ in (\ref{eq:genintz}) falls faster
than $(-\eta)^4$ as $\eta \rightarrow 0^{-}$. This is faster than the
IR contribution which falls only as $(-\eta)^{\beta_i}$ if $\beta_i
\le 4$. If $\beta_i > 4$ then both the IR and UV contributions fall
faster than $(-\eta)^4$. Hence, the $S_i(z)$ for which $\beta_i > 4$
are subdominant at late times.  Thus we conclude that the leading
order state dependent terms at late times are those with the smallest
$\beta_i$.  These give an IR dominant, ({\it i.e.} finite $k$, $z\ll
1$) contribution of the form, 
\begin{equation} I(\eta) \rightarrow
(-\eta)^{\beta_i} \int_0^{\infty}\, {\rm d}k \ k^{\beta_i -1}\ R(k)
\rightarrow (-\eta)^{3- 2\nu} \int_0^{\infty}\, {\rm d}k \ k^{2-
2\nu}\ R(k)\,, 
\label{eq:Ilim} 
\end{equation} as $\eta\rightarrow
0^{-}$. This last integral is guaranteed to converge at its upper
limit for all $\Re (\nu) \ge -{1\over 2}$, and in particular for $0
\le \Re (\nu) \le {3\over 2}$ by (\ref{eq:adbR}). Hence for all $\Re
(\nu) < {3\over 2}$ the state dependent contributions go to zero as
$a^{2\nu -3}$ at late times. The limiting case when $\nu \rightarrow
{3\over 2}$ is special because then the state dependent IR
contributions apparently does not go to zero at late times. This case
will be considered separately in the next Section.

We will now make the proof more explicit by giving the form of all of
the state dependent terms of the scalar field in de Sitter space, and
analyzing the IR and UV contributions in detail. If we make use of
Eq. (\ref{eq:c1c2K0}), we find that in an arbitrary state
\begin{eqnarray}
\langle T_{ab} \rangle_{ren} &=& \langle T_{ab}
\rangle_{BD} + \langle T_{ab} \rangle_{SD} \;,
\end{eqnarray}
where $\langle T_{ab}\rangle_{SD}$ is the finite state dependent term,
depending on the coefficients $c_1(k), c_2(k)$, and $n_k$, which may
be expressed as an integral over the wave number $k$ in the form
\begin{eqnarray}
\langle T_{ab} \rangle_{SD} &=& \frac{1}{4 \pi^2} \int_0^{\infty }
{\mathrm{d}} k
 \; I_{a b}(k,\eta) \; .    \label{eq:Iab}
\end{eqnarray}
The explicit expressions for the integrand $I_{ab}$ depend on whether
$\nu$ is real and not an integer, real and an integer, or pure
imaginary, although the result will be the same for all $\Re (\nu)<
{3\over 2}$. In the rest of this Section we restrict our discussion to
the case where $\nu$ is real and not equal to an integer. The cases of
integer and imaginary values of $\nu$ are covered in Appendix A.  For
real $\nu$ after some regrouping of terms we have
\begin{mathletters} 
\begin{eqnarray}
{I^0}_0 &=& A_1(k) \left[S_1 + \frac{1}{4} (9 + 4 m^2 - 48 \xi) 
	S_4 - \frac {3}{2} (4 \xi -1) S_7 + S_{10} \right] \nonumber \\
         &  &  + A_2(k) \left[S_2 + \frac{1}{4} (9 + 4 m^2 - 48 \xi) S_5 - 
            \frac{3}{2} (4 \xi -1) S_8 + S_{11} \right] \nonumber \\
         &  &  + A_3(k) \left[S_3 + \frac{1}{4} (9 + 4 m^2 - 48 \xi) S_6 - 
            \frac{3}{2} (4 \xi -1) S_9 + S_{12} \right] 
\; ,\qquad {\rm and} \label{eq:I00} \\
I &=& A_1(k) \left[-2(6 \xi - 1) S_1 - \frac{1}{2}\left(9 + 8 (3 \xi - 1) m^2 
      - 102 \xi + 288 \xi^2\right)S_4 + 3(6 \xi - 1) S_7 + 2(6 \xi -1) S_{10}
	 \right] \nonumber \\ 
  &  & + A_2(k) \left[-2(6 \xi - 1) S_2 
- \frac{1}{2}\left(9 + 8 (3 \xi - 1) m^2  
      - 102 \xi + 288 \xi^2\right)S_5 + 3(6 \xi - 1) S_8 + 2(6 \xi -1) S_{11} 
	\right] \nonumber \\ 
  &  & + A_3(k) \left[-2(6 \xi - 1) S_3 - \frac{1}{2}\left(9 
+ 8 (3 \xi - 1) m^2  
      - 102 \xi + 288 \xi^2\right)S_6 + 3(6 \xi - 1) S_9 + 2(6 \xi -1) S_{12}
	 \right] 
\; ,
\label{eq:I}
\end{eqnarray}
\end{mathletters}
with
\begin{mathletters}
\begin{eqnarray}
A_1(k) &=& -\frac{\pi}{2 k} \left[\csc^2(\nu \pi) \left((1+ 2 n_k) |c_2|^2 + 
	n_k\right) + \frac{1}{2} \left(1-\cot^2(\nu\pi)\right)(1+2 n_k) 
	(c_1 c_2^* + c_1^* c_2)\right. \nonumber \\
     &  & \left. + \frac{i}{2} \cot(\nu\pi) (1+2 n_k)(c_1 c_2^* - c_1^* c_2)
	\right] 
\; ,\\ 
A_2(k) &=& \frac{\pi}{k} \left[\cot(\nu \pi) \csc(\nu\pi) \left((1+ 2 n_k) 
	|c_2|^2 + n_k\right) - \frac{1}{2} \cot(\nu\pi)\csc(\nu\pi)(1+2 n_k) 
	(c_1 c_2^* + c_1^* c_2) \right.\nonumber \\
     &  & \left. + \frac{i}{2} \csc(\nu\pi) (1+2 n_k) (c_1 c_2^* - c_1^* c_2)
	\right] 
\; , \\ 
A_3(k) &=& -\frac{\pi}{2k} \left[\csc^2(\nu\pi) \left((1+ 2 n_k) |c_2|^2 + n_k
	\right) - \frac{1}{2} \csc^2(\nu\pi)(1+2 n_k) 
(c_1 c_2^* + c_1^* c_2) \right]\, ,
\label{eq:Adef}
\end{eqnarray}
\end{mathletters}%
and the $S_i(-k\eta)$ composed of various products of $J_\nu(-k\eta)$,
$J_{-\nu}(-k\eta)$ and their derivatives.  Making use of the general
formula for the product of two Bessel functions~\cite{gradrhyz},
\begin{equation}
J_\mu(z) J_\nu(z) = \sum_{p=0}^{\infty}\frac{(-1)^p 
\left(\frac{z}{2}\right)^{\nu+\mu+2p}\Gamma(\nu+\mu+2p+1) }
{p! \Gamma(\nu+\mu+p+1) \Gamma(\nu+p+1) \Gamma(\mu+p+1)} \,,
\label{eq:JJ}
\end{equation}
we can expand the $S_i$ in power series of the form,
\begin{equation}
S_i(z) = \sum_{p=0}^{\infty} s_{i,p}\ z^{2p + \beta_i} \;, \label{eq:Si}
\end{equation}
with $z = -k\eta$.
The explicit expressions for $\beta_i$ and $S_i$ for the case of real,
non-integer $\nu$ are given in Table 1.
\hfil\break
\vspace{.4cm}
\begin{center}
\begin{tabular}{|c||c|c|} \hline 
\rule[-0.3cm]{0mm}{0.80cm}
 $\;\;i\; \; $ &$\; \; \; \; \beta_i \; \; \; \; $&$\; \;  S_i(z)\; \; \;$
 \\ 
\hline
\hline
\rule[-0.3cm]{0mm}{0.80cm}
 1 & $5 + 2 \nu$ &  $z^5 J_\nu^2(z)$ \\ 
\hline 
\rule[-0.3cm]{0mm}{0.80cm}
 2 & $5     $ &  $z^5 J_\nu(z) J_{-\nu}(z)$ \\ \hline
\rule[-0.3cm]{0mm}{0.80cm}
 3 & $5 - 2 \nu$ &  $z^5 J_{-\nu}^2(z)$ \\ \hline
\rule[-0.3cm]{0mm}{0.80cm}
 4 & $3 + 2 \nu$ &  $z^3 J_\nu^2(z)$ \\ \hline
\rule[-0.3cm]{0mm}{0.80cm}
 5 & $3     $ &  $z^3 J_\nu(z) J_{-\nu}(z)$ \\ \hline
\rule[-0.3cm]{0mm}{0.80cm}
 6 & $3 - 2 \nu$ &  $z^3 J_{-\nu}^2(z)$ \\ \hline
\rule[-0.3cm]{0mm}{0.80cm}
 7 & $3 + 2 \nu$ &  $z^4 \frac{\rm d}{{\rm d}z} J_\nu^2(z)$ \\ \hline
\rule[-0.3cm]{0mm}{0.80cm}
 8 & $3     $ &  $z^4 \frac{\rm d}{{\rm d}z} 
\left(J_\nu(z) J_{-\nu}(z)\right)$ \\ \hline
\rule[-0.3cm]{0mm}{0.80cm}
 9 & $3 - 2 \nu$ &  $z^4 \frac{\rm d}{{\rm d}z} J_{-\nu}^2(z)$ \\ \hline
\rule[-0.3cm]{0mm}{0.80cm}
 10 & $3 + 2 \nu$ &  $z^5 \left(\frac{\rm d}{{\rm d}z} J_\nu (z)\right)^2$ 
\\ \hline
\rule[-0.3cm]{0mm}{0.80cm}
 11 & $3     $ &  $z^5 \left(\frac{\rm d}{{\rm d}z} J_\nu(z)\right)
                      \left(\frac{\rm d}{{\rm d}z} J_{-\nu}(z)\right)$ 
\\ \hline
\rule[-0.3cm]{0mm}{0.80cm}
 12 & $3 - 2 \nu$ &  $z^5 \left(\frac{\rm d}{{\rm d}z} J_{-\nu}(z)\right)^2$ 
\\ \hline
\end{tabular}
\end{center}
\begin{center}
{Table 1}
\end{center}

We are interested in the behavior of the finite state dependent terms,
$\langle T_{ab} \rangle_{SD}$, in the limit $\eta \rightarrow 0^- $.  
To investigate this limit in detail it is useful to break up the integral in 
Eq.\ (\ref{eq:Iab}) into the three parts
\begin{equation}
\langle T_{ab} \rangle_{SD}
=  \frac{1}{4 \pi^2} \int_0^\lambda
{\mathrm{d}} k \; I_{ab}(k,\eta)
+ \frac{1}{4 \pi^2}\int_\lambda^{-Z/\eta}  {\mathrm{d}} k\; I_{ab}(k,\eta)
+ \frac{1}{4 \pi^2}\int_{-Z/\eta}^{\infty} {\mathrm{d}} k\;
  I_{ab}(k,\eta)
\; .
\label{eq:thrI}
\end{equation}
Here $\lambda$ is a finite positive constant.  For $k>\lambda$ we can
make use of the UV conditions (\ref{eq:ckbound}), (\ref{eq:nkbound}),
and (\ref{eq:CNlim}) in the second and third integrals.  Thus, the
most infrared sensitive integral is the first one.  The positive
constant $Z$ is arbitrary, provided only that $Z > -\lambda\eta$,
which is always satisfied for fixed $\lambda$ and small enough
$-\eta$. Hence the second integral provides the bulk of the
contribution of the state dependent wave numbers that have redshifted
outside the de Sitter horizon at late times. The third integral is the
contribution of the state dependent terms still within the horizon
which go to zero very rapidly at late times due to the UV
conditions. If the $\eta \rightarrow 0^-$ limit is uniform then all
three integrals should vanish unconditionally in this limit, {\it
i.e.} without any restrictions on the arbitrary parameters $\lambda$
and $Z$.

We begin by analyzing the first integral in (\ref{eq:thrI}).  We are
considering only fourth order adiabatic states that are IR
admissible. Hence the $k$ integration converges at its lower limit and
we can expand the integrand for this first integral in a series of the
form (\ref{eq:Si}) and interchange the order of summation and
integration. Each term in the resulting sums contains an integral over
$k$ that is finite by assumption and a factor of $(-\eta)^{2 p +
\beta_i}$ where $p =0, 1, 2\dots$ is a non-negative integer. In fact,
since at late times, $-\lambda \eta \ll 1$ for any finite $\lambda$,
it is sufficient to consider only the leading $p=0$ term. From Table 1
it is clear that for $\nu < {3\over 2}$, $\beta_i > 0$ for all
$i$. Hence in the limit $\eta \rightarrow 0^-$, the integrand vanishes
and therefore, the first integral on the right hand side of
Eq. (\ref{eq:thrI}) vanishes in this late time limit.

For the second integral in (\ref{eq:thrI}) we may utilize the
expansion (\ref{eq:JJ}) again.  Since
the integration is between finite limits for finite $\eta$ we can
exchange the order of summation over $p$ and integration over
$k$. Having done so we can then bound each term in the sums by taking
the absolute value of its factors. Inspection of the expressions
(\ref{eq:I}) and (\ref{eq:Adef}) indicates that the result is a linear
combination of terms involving integrals of the three possible forms
\begin{eqnarray}
{\cal I}_1&=& \int_\lambda^{-Z/\eta} \frac{{\rm d}k}{k} \ |c_2(k)|^2\ 
(1 + 2 n_k)
\,(-k\eta)^{2 p + \beta_i}\,, \nonumber \\
{\cal I}_2&=& \int_\lambda^{-Z/\eta} \frac{{\rm d}k}{k}\ 
|c_1(k) c_2(k)|\ (1 + 2n_k)
\,(-k\eta)^{2 p + \beta_i} \,, \nonumber \\
{\cal I}_3&=& \int_\lambda^{-Z/\eta} \frac{{\rm d}k}{k}\ n_k 
\,(-k\eta)^{2 p + \beta_i}
\; ,
\label{eq:J12}
\end{eqnarray}
multiplied by constant coefficients.  Because of (\ref{eq:CNlim})
for any $\lambda$
\begin{eqnarray}
\vert C(k)\vert &<& C\,,\nonumber\\
N(k)&<& N\,,
\end{eqnarray}
for some real positive numbers $C$ and $N$, depending on $\lambda$.
With these bounds we can bound the values of $|c_2(k)|$ and $n_k$, and
use Eq.\ (\ref{eq:c1c2K0}) to bound $|c_1(k)|$  for all $k \ge \lambda$
as follows
\begin{eqnarray}
 |c_2(k)| &<&  \frac{C}{k^4}\; , \nonumber \\
 1 \le |c_1(k)| &=& \left[1 + \vert c_2(k)\vert^2\right]^{1\over 2}
\le 1 + \vert c_2(k)\vert^2 < 1 + {C^2\over k^8} \le 1 + \frac{C^2}{\lambda^8}\; ,\nonumber  \\
  n_k &<& \frac{N}{k^4} \nonumber \\
 1 + 2 n_k &<& 1 + \frac{2N}{k^4} \le 1 + \frac{2 N}{\lambda^4} \; .
\label{eq:c1c2nkbounds} \end{eqnarray} 
Using these bounds the integrals in Eq.\ (\ref{eq:J12}) can be bounded
as follows
\begin{eqnarray}
{\cal I}_1&< & C^2 (1 + 2N/\lambda^4) \int_\lambda^{-Z/\eta} \,{\rm d}
k\,k^{2p +\beta_i - 9}
 (-\eta)^{2p + \beta_i}\nonumber\\
&=& \frac{C^2(1 + 2N/\lambda^4)}{\beta_i + 2p - 8} \left[(-\eta)^8 Z^{2p + \beta_i - 8} 
- (-\eta)^{2p + \beta_i} \lambda^{2p + \beta_i - 8} \right] \nonumber \\
{\cal I}_2&< & C(1+C^2/\lambda^8)(1+ 2N/\lambda^4) \int_\lambda^{-Z/\eta} \, {\rm d}k\,
k^{2p +\beta_i - 5} (-\eta)^{2p + \beta_i}\nonumber\\
&=& \frac{ C(1+C^2/\lambda^8)(1+ 2N/\lambda^4)}{\beta_i + 2p - 4} 
\left[(-\eta)^4 Z^{2p + \beta_i - 4} - (-\eta)^{2p + \beta_i} 
\lambda^{2p + \beta_i - 4}\right]\nonumber \\
{\cal I}_3&< & N \int_\lambda^{-Z/\eta} \,{\rm d}k\,k^{2p +\beta_i - 5}
 (-\eta)^{2p + \beta_i}\nonumber\\
&=& \frac{N}{\beta_i + 2p - 4} \left[
(-\eta)^4 Z^{2p + \beta_i - 4}- 
\lambda^{2p + (-\eta)^{2p + \beta_i} \beta_i - 4}\right]\,.
\label{eq:J12value}
\end{eqnarray}
Each of these bounds vanishes in the limit $\eta \rightarrow 0^-$.
Therefore each of the terms appearing in the second integral in
(\ref{eq:thrI}) vanishes in the late time limit for $\Re (\nu) <{3\over 2}$.

If $\nu = {1\over 2}$ some of the terms will have vanishing denominators and
should be interpreted according to the limiting relation
\begin{equation}
\lim_{q \rightarrow 0} { \left(-{Z\over \eta}\right)^q - \lambda^q \over q} =
\log \left({Z\over -\eta\lambda}\right) \,,
\end{equation}
but the appearance of these logarithms does not change the result
since they are always multiplied by at least $(-\eta)^{2p +\beta_i}$
which vanishes for $\Re (\nu) <{3\over 2}$ for $p = 0, 1, 2\dots$.
The cases $\nu = 0, 1$ also involve logarithms in the Bessel function
expansions but the result that the second integral in (\ref{eq:thrI})
vanishes in the late time limit is unchanged.  One may consider also
the case when $\nu = i \gamma$ is pure imaginary, where the forms of
the coefficients (\ref{eq:Adef}) and bilinears $S_i(z)$ in the table
change somewhat, with again the same result. For completeness these
cases are treated in detail in Appendix A.

Finally, for the third integral on the right hand side of Eq.\
(\ref{eq:thrI}) we do not expand the Bessel functions in powers of
$(-k\eta)$.  Instead we change the integration variable to $z = -k
\eta$ and find integrals of the forms
\begin{eqnarray}
  {\cal J}_1 &=& (-\eta)^8 \int_Z^{\infty} \frac{{\rm d}z}
{z^9}\ \Big|C\left(\frac{z}{-\eta}\right)\Big|^2\ 
  \left[ 1 + 2\left({-\eta\over z}\right)^4 N
\left(\frac{z}{-\eta}\right)\right] \  
 S_i(z)\,, \nonumber \\
 {\cal J}_2 &=& (-\eta)^4 \int_Z^{\infty} \frac{{\rm d}z}
{z^5}\ (\Re\ {\rm or}\ \Im)
 \left\{c_1\left({z\over -\eta}\right) \left[C^*\left(\frac{z}{-\eta}\right)
\right]
\right\}\ 
 \left[1 + 2\left({-\eta\over z}\right)^4 
N\left(\frac{z}{-\eta}\right) \right]\ S_i(z)\,, \nonumber \\
 {\cal J}_3 &=& (-\eta)^4 \int_Z^{\infty} \frac{{\rm d}z}
{z^5}\ N\left(\frac{z}{-\eta}\right) S_i(z)\,.
\end{eqnarray}
Since we are considering fourth order adiabatic states these integrals
are finite for all $-\infty <\eta \le 0$.  Furthermore, the integrands
of these integrals are finite throughout the entire integration range
for $-\infty < \eta \le 0$. With the lower limit $Z > 0$ fixed, it is
clear that we can evaluate the integrals at $\eta = 0$ by first taking
the limit $\eta \rightarrow 0^-$ of the integrands and then computing
the integrals.  Since the integrands all vanish in the limit $\eta
\rightarrow 0^-$, the integrals do as well. Then the third integral in
(\ref{eq:thrI}) vanishes in the late time limit for $\Re (\nu)
<{3\over 2}$. From the adiabatic four conditions (\ref{eq:CNlim})
these UV contributions go to zero faster than $(-\eta)^4$ for any
fourth order adiabatic state.

Therefore we have proven that for $\Re (\nu) < {3\over 2}$ all the
integrals in Eq.\ (\ref{eq:thrI}) vanish in the limit $\eta
\rightarrow 0^-$ for an arbitrary state that is both infrared and
ultraviolet finite. The fact that all three integrals vanish at late
times, {\it independently} of the parameters $\lambda$ and $Z$ which
we introduced to control the IR and UV contributions of the mode
integral verifies that the total mode integral does converge uniformly
in $\eta$, as expected. Hence all the state dependent contributions
vanish at late times, $\eta \rightarrow 0^-$, and we have proven that
in this limit $\langle T_{a b} \rangle_{ren} \rightarrow \langle T_{a
b} \rangle_{BD}$. Thus, we conclude that for $\Re(\nu)<{3\over 2}$ the
value of the energy-momentum tensor in any physically admissable
homogeneous and isotropic RW quantum state asymptotically approaches
the Bunch-Davies value in de Sitter space at late times.

Moreover, from inspection of (\ref{eq:J12value}) we observe that the
contribution from the upper limit at $Z$ in the second integral always
falls faster than $(-\eta)^4$, while the contribution from the lower
limit at $\lambda$ falls off only as $(-\eta)^{\beta_i}$ as $\eta
\rightarrow 0^-$. This must be the same order as the first IR integral
so that the arbitrary parameter $\lambda$ drops out of the final
result. Hence our detailed evaluation has verified that the leading
order state dependent corrections to the BD expectation value at late
times come from the terms in $\langle T_{a b} \rangle_{ren}$ with the
smallest $\beta_i$. From Table 1 these are the $i=6, 9$, and $12$
terms. Collecting these terms from (\ref{eq:I}) and (\ref{eq:Adef}),
and the numerical coefficients from the Bessel function product
formula (\ref{eq:JJ}), we conclude that the leading order behavior of
the energy density and trace at late times are given by
\begin{mathletters}
\begin{eqnarray}
\varepsilon &\rightarrow &  a^{2\nu -3} \left[\left(\nu -{1\over 2}\right)^2 
+ m^2 + 2(6\xi -1)(\nu -1)\right] \int_0^{\infty}\ {\rm d}k\ k^{2-2\nu} R(k)
\,, \qquad {\rm and}\label{eq:lateps}\\
T &\rightarrow& 2 a^{2\nu -3} \left\{-m^2 + (6\xi-1) \left[-\left(\nu 
-{1\over 2}\right)^2 + 2\nu + m^2 + 2(6\xi -1)\right]\right\} 
\int_0^{\infty}\ {\rm d}k\ k^{2-2\nu} R(k)\,,
\label{eq:lateT}
\end{eqnarray}
\end{mathletters}
respectively, with
\begin{equation}
R(k) = {2^{2\nu - 3}\csc^2 (\pi\nu)\over \pi \left[\Gamma (-\nu + 1)\right]^2} 
\left\{ (1+ 2 n_k) \left[|c_2(k)|^2 - \Re (c_1 c_2^*)\right] + n_k \right\} \,.
\end{equation}
The leading order state dependent contribution at late times is $a^{2
\nu -3}$ from the IR part of the mode integral, which falls off very
slowly if $\Re (\nu) \rightarrow {3\over 2}$. We examine this latter
limit in detail in the next Section.

\section{The Allen-Folacci Attractor for $\Re (\nu) \rightarrow {3\over 2}$}
\label{sec:nu32}

In the analysis of the previous Section we saw that the terms with the
slowest falloff at late times were those with the smallest $\beta_i =
3 - 2 \nu$ which behave like $a^{2 \nu - 3}$ for $\nu$ real and
positive, and the coefficient of this falloff is controlled by the
finite $k$, IR part of the mode integral. To examine the limit $\nu
\rightarrow {3\over 2}$ carefully, it is easiest to work with closed
spatial sections and a discrete set of mode functions in order to
treat the most infrared sensitive, spatially homogeneous $k=1$ mode
separately from the rest, instead of dealing with an infrared
sensitive continuous mode integral.

The scale factor for $\kappa = +1$ is given by Eq.\ (\ref{eq:ak1}) and
under the variable substitution $\zeta= i \tan \eta$ the mode equation
(\ref{eq:mode}) becomes Legendre's differential equation. Hence the 
general solutions may be expressed in terms of associated Legendre 
functions $P^{\pm k}_{-{1\over 2} + \nu}(\zeta)$. Since as conventionally 
defined these functions have a cut discontinuity on the real axis from 
$-1$ to $1$ if $k$ is an odd integer, we write the fundamental complex 
valued solution for real $\nu$ in the form \begin{eqnarray} f_k(\eta) &=&
\left[{\Gamma\left(k + \frac{1}{2} + \nu\right) \Gamma\left(k +
\frac{1}{2} - \nu\right)\over 2}\right]^{1\over 2} \exp
\left(-{ik\pi\over 2} \epsilon (\eta)\right) P^{-k}_{-{1\over 2} +
\nu} (i \tan \eta) \nonumber\\ &=& \left[{\Gamma\left(k + \frac{1}{2}
+ \nu\right) \Gamma\left(k + \frac{1}{2} - \nu\right)\over
2}\right]^{1\over 2} {e^{-ik\eta}\over k!}\,F\left(\frac{1}{2} + \nu ,
\frac{1}{2} - \nu ;k+1; {1 -i\tan \eta\over 2}\right)\,,
\label{eq:hyper} \end{eqnarray} where $\epsilon (\eta) = \theta(\eta)
- \theta(-\eta)= \pm 1$ is the sign function and $F$ is the
hypergeometric function. The phase factor depending on the sign of
$\eta$ for odd $k$ removes the discontinuity in the $P^{-k}_{-{1\over
2} + \nu}$ function as $\eta$ approaches zero from positive or
negative values, as the second form of (\ref{eq:hyper}) makes clear,
since the $F$ function is an analytic function of $1-\zeta\over 2$,
with no branch cuts for $\zeta$ on the imaginary axis. With the
normalization factors chosen as in (\ref{eq:hyper}) the Wronskian
condition \begin{equation} f_k {f_k^*}' - f_k^* f_k' =i\,,
\end{equation} is satisfied. Hence the general solution of the mode
equation (\ref{eq:mode}) is \begin{equation} \psi_k(\eta) = \alpha_k
f_k(\eta) + \beta_k f_k^*(\eta)\,, \label{eq:psik} \end{equation} with
\begin{equation} |\alpha_k|^2 - |\beta_k|^2 = 1\,.  \label{eq:AkBk}
\end{equation} The Bunch-Davies state is given by $\alpha_k = 1$ and
$\beta_k = 0$.

Now as $\nu \rightarrow {3\over 2}$ inspection of (\ref{eq:hyper})
shows that all the $f_k$ for $k>1$ are regular.  In fact, in that
limit the hypergeometric series for $F$ terminates and the $f_k
(\eta)$ become the elementary functions
\begin{equation}
\lim_{\nu \rightarrow {3\over 2}}f_k(\eta) =
\frac{e^{-i k \eta}}{[2 k(k^2 - 1)]^{1\over 2}}\, (k + i\tan \eta)\,,\qquad
k = 2, 3, \dots
\; ,
\label{eq:psiknon}
\end{equation}
so they can be treated in the Bunch-Davies state $\alpha_k = 1,
\beta_k = 0$ with no difficulty. However in this limit the $k=1$ mode
function is singular and must be treated separately. The behavior of
the $k=1$ mode as $\nu \rightarrow {3\over 2}$ is similar to that of a
simple harmonic oscillator mode as its frequency goes to zero, {\it
i.e.} in the limit where the harmonic oscillator becomes a free
particle.  The zero frequency limit in this simple flat space analogy
is reviewed in Appendix B.  Just as in that case, one can construct
regular solutions to the mode equation in the limit $\nu \rightarrow
{3\over 2}$ by taking suitable linear combinations of $f_1$ and
$f_1^*$. In fact, the limiting form of $f_1 (\eta)$ can be found from
Sec. 2.3.1 of Ref. \cite{Bate}, which gives
\begin{equation}
f_1(\eta) \rightarrow {\sec \eta \over 2 \sqrt{{3\over 2} - \nu}}
-{i\over 2} \sqrt{{3\over 2} - \nu}\ \sec \eta\ (\eta + \sin\eta\,\cos\eta) + 
\dots
\; .
\end{equation}
We have neglected terms in the real part of $f_1$ that are of order
$\sqrt{{3\over 2} - \nu}$.  We have also neglected all terms that go
to zero faster than this as $\nu \rightarrow {3\over 2}$. Extracting
the scale factor $a(\eta) = \sec\eta$ we now define the real functions
$u$ and $v$ by
\begin{eqnarray}
u(\eta)&\equiv &{1\over a(\eta)}\lim_{\nu \rightarrow 
{3\over 2}} \left\{\sqrt{{3\over 2} - \nu}
\ (f + f^*)\right\} = 1\,,\nonumber\\
v(\eta) &\equiv & {i\over a(\eta)}\ \lim_{\nu \rightarrow 
{3\over 2}} \left\{{1\over 2\sqrt{{3\over 2} - \nu}}\
(f - f^*)\right\} =
{\eta + \sin\eta\,\cos\eta\over 2}  \,.
\label{uvdeS}
\end{eqnarray}

We next define new coefficients
\begin{eqnarray}
A &\equiv& -i\ \lim_{\nu \rightarrow {3\over 2}}
\left\{ \sqrt{{3\over 2} - \nu}\ (\alpha_1 - \beta_1)
\right\}
\; ,
\nonumber\\
B &\equiv& \lim_{\nu \rightarrow {3\over 2}}
\left\{ {1\over 2\sqrt{{3\over 2} - \nu}}\ (\alpha_1 + \beta_1)
\right\}
\; ,
\label{ABdeS}
\end{eqnarray}
which are finite in this limit. With these definitions the
normalization is such that
\begin{equation}
A^*B - B^*A = i
\; ,
\end{equation}
and the limit of the general $k=1$ mode function may be written
\begin{equation}
\lim_{\nu \rightarrow {3\over 2}}\psi_1(\eta) = \sec\eta\ (Av + Bu) 
= a(\eta) \left[ {A\over 2} \ (\eta + \sin\eta\,\cos\eta)+ B\right]\,.
\label{eq:psionelim}
\end{equation}
One can also define the time-independent Hermitian operators,
\begin{eqnarray}
Q &\equiv &  {1\over 2\sqrt{{3\over 2} - \nu }}
\left[(\alpha_1 + \beta_1)\ a_1 +
(\alpha_1 + \beta_1)^*\ a_1^{\dagger}\right] \rightarrow 
Ba_1 + B^* a_1^{\dagger}
\,,\nonumber\\
P &\equiv & -i \sqrt{{3\over 2} - \nu}\ \left[(\alpha_1 - \beta_1)\ a_1
- (\alpha_1 - \beta_1)^*\ a_1^{\dagger}\right] 
\rightarrow Aa_1 + A^* a_1^{\dagger}\,,
\label{eq:QPdef}
\end{eqnarray}
obeying the canonical commutation relations, $[Q, P] = i$.  They 
also remain finite in the limit $\nu \rightarrow {3\over 2}$.

We will consider the limit $\nu \rightarrow {3\over 2}$ of the energy
density in the particular order of fixing the mass of the scalar field
to be $m=0$ and letting the dimensionless parameter $\xi$ approach
zero, since this case is relevant for the infrared scaling analysis of
Section VI. From (\ref{eq:nu}) with $m=0$ and $\xi$ small we have
\begin{equation}
\xi \simeq {(3 - 2 \nu)\over 8} \rightarrow 0
\;.
\label{eq:xizero} 
\end{equation}
The complementary case, $\xi =0$ and $m^2 \rightarrow 0$ is similar
and has been discussed in Refs. \cite{a-f} and \cite{k-g}. From
(\ref{eq:T00u}) we read the energy density in the $k=1$ mode for $m=0$
\begin{equation}
\varepsilon_1 = -\langle {T^0}_0\rangle_{k=1} =
{(1 + 2n_1)\over 4\pi^2 a^4}\left\{ |\psi_1'|^2 + |\psi_1|^2
+ (6 \xi -1) [\tan\eta\ (\psi_1\psi_1^{*\prime} + \psi_1^*\psi_1')
- (\tan^2\eta - 1)|\psi_1|^2]\right\}\,.
\end{equation}
We are interested first in the asymptotic form of this energy density 
at late times, $\eta \rightarrow {\pi \over 2}$, and then in the limiting
form of the resulting expression as $\xi \rightarrow 0$ according
to (\ref{eq:xizero}). The asymptotic late time limit of the
mode function $f_1 (\eta)$ for any $\nu > 0$ can be found from
the inversion transformation of the hypergeometric function,
given by formula 2.1.4 (17) of \cite{Bate}. We find that
\begin{eqnarray}
f_1 \left(\eta \rightarrow {\pi\over 2}\right) &\rightarrow &
\left[ {\Gamma \left( {3\over 2} - \nu\right) \over 2 \Gamma \left( {3\over 2}
+ \nu\right)}\right]^{1\over 2} {\Gamma (2 \nu) \over \Gamma 
({1\over 2} + \nu)}
(-i)\left({i\sec \eta\over 2}\right)^{\nu - {1\over 2}} 
\sim a^{\nu - {1\over 2}}\, ,\\
 f_1' &\rightarrow & \left(\nu - {1\over 2}\right)\ \sec\eta\ f_1 \sim 
a^{\nu + {1\over 2}}\, ,  
\end{eqnarray}
as $ \eta \rightarrow {\pi\over2}$, which is the same late time
behavior in terms of the scale factor that we found in the spatially
flat coordinates.  Notice that at $\nu = {3\over 2}$ the phase factor
cancels and these limiting forms of the oscillatory mode functions
become real, while the $\Gamma$ function has a pole singularity
there. Thus, keeping only the leading behavior as $\nu$ approaches
$3\over 2$ we find that the late time limits of the mode function and
its derivative are
\begin{eqnarray}
\psi_1 \left(\eta \rightarrow {\pi\over 2}\right) &\rightarrow &
{\alpha_1 + \beta_1 \over 2\sqrt{{3\over 2} - \nu}}\ a^{\nu - {1\over 2}}\,,
\nonumber\\
\psi_1' \left(\eta \rightarrow {\pi\over 2}\right) &\rightarrow &
{\alpha_1 + \beta_1 \over 2\sqrt{{3\over 2} - \nu}}\ a^{\nu + {1\over 2}}\,.
\label{eq:psikone}
\end{eqnarray}
Thus, the dominant terms in the $k=1$ energy density $\varepsilon_1$
in this limit are $|\psi_1'|^2 $ and the terms involving
${a'\over a} = \tan \eta$, and we obtain
\begin{eqnarray}
\lim_{\eta \rightarrow {\pi \over 2}} \varepsilon_1 = 
\left[{3(1 + 2 n_1)\over 32\pi^2 }|\alpha_1 + \beta_1|^2 + 
{\cal O}(3 - 2\nu)\right] a^{2\nu -3}\,.
\label{eq:epsone}
\end{eqnarray}
The singularity in the $\Gamma$ function has canceled against the $\xi$ in
the numerator. The remaining coefficient is clearly state dependent. From this
asymptotic form of the energy density in the $k=1$ mode at late times
it is clear that for $\nu$ close to but still less than $3\over 2$,
the state-dependent energy density $\varepsilon_1$ goes to zero at
late times, albeit very slowly, which is consistent with our previous
flat section analysis.

The corresponding expression for the trace in the $k=1$ mode is
\begin{equation}
-\varepsilon_1 + 3p_1 =
{(1 + 2n_1)\over 2\pi^2 a^4} (6 \xi -1)\left\{-|\psi_1'|^2 
+  [\tan\eta\ (\psi_1\psi_1^{*\prime} + \psi_1^*\psi_1')
+ [\sec^2\eta + 2(6 \xi -1)\sec^2\eta ]|\psi_1|^2]\right\}\,.
\end{equation}
Substituting the late time asymptotic forms (\ref{eq:psikone}) 
to leading order in $\xi$ as before yields
\begin{equation}
-\varepsilon_1 + 3p_1 \rightarrow  \left[-{3(1 + 2n_1)\over 8\pi^2 }
|\alpha_1 + \beta_1|^2 +  {\cal O}(3 - 2\nu)\right] a^{2\nu -3} 
\rightarrow -4\varepsilon_1\, .
\label{eq:trone}
\end{equation}
Hence, $p_1 \rightarrow -\varepsilon_1$ and the contribution from this
mode is de Sitter invariant at late times for any initial physical
state.

Since the Bunch-Davies state is $\alpha_k =1$ and $\beta_k = n_k = 0$
for all $k$ and this state has a renormalized $\langle T_{ab} \rangle$
which is strictly time independent for all $\nu$, the time dependent
contribution in (\ref{eq:epsone}) and (\ref{eq:trone}) for $\nu <
{3\over 2}$ in the $k=1$ mode must be canceled by a time dependent
contribution from all the other modes in the renormalized
energy-momentum tensor. In other words, the late time behavior of all
the $k>1$ modes in the BD state must be
\begin{equation}
\varepsilon_{BD}\Big\vert_{k >1} \rightarrow \varepsilon_{BD} - 
\left[{3\over 32\pi^2 } + {\cal O}(3 - 2\nu)\right] a^{2\nu -3} \; ,
\label{eq:kgone}
\end{equation}
for $\nu$ close to $3\over 2$. The pressure for the $k>1$ modes is obtained
from this by the de Sitter invariant relation $p=-\varepsilon$.

The subtraction of the second term in (\ref{eq:kgone}) can be understood in
a different way. Consider the short distance expansion of 
the BD two-point function for $\nu < {3\over 2}$, namely~\cite{k-g}
\begin{equation}
G_{BD}(x,x') \rightarrow {1\over 8\pi^2}\left[ {1\over 1 - Z} - \log (1-Z)
+ {1\over {3\over 2} - \nu}\right]\,,
\label{BDG}
\end{equation} 
for the de Sitter invariant bi-scalar, $Z(x,x') \rightarrow 1$ and
$\nu$ close to $3\over 2$. The constant term is singular at $\nu =
{3\over 2}$, but it gives a finite contribution to the energy density
from the $\xi G_{ab} \phi^2$ term in the energy-momentum tensor.  In the
dimensionless units we are using this is equal to
\begin{equation}
{3\over 8\pi^2}\ \xi\ \left({R\over 12}\right)^2{1\over {3\over 2} - \nu} = 
{3\over 32\pi^2}\,,
\label{eq:BDUV}
\end{equation}
for $\xi \rightarrow 0$ according to (\ref{eq:xizero}).  The last
constant term in (\ref{BDG}) is absent in the short distance expansion
of the Allen-Folacci two-point function~\cite{a-f}, and in the
Bunch-Davies two-point function it comes entirely from the $k=1$ mode
in the BD state. Hence the contribution to the energy density of the
$k>1$ modes in the BD state does not contain (\ref{eq:BDUV}), and must
equal the full $\varepsilon_{BD}$ minus (\ref{eq:BDUV}) from the $k=1$
mode, which is equivalent to (\ref{eq:kgone}) at $\nu = {3\over 2}$.

With the $k=1$ and $k>1$ mode contributions separated we are now in a
position to take the limit of $\nu \rightarrow {3\over 2}$.
Apparently we would obtain a state dependent contribution from
(\ref{eq:epsone}).  However, inspection of (\ref{eq:psionelim}) shows that
requiring the $k=1$ mode function to be regular as $\nu \rightarrow
{3\over 2}$ is equivalent to requiring that $A$ and $B$ remain finite in this
limit. From (\ref{ABdeS}) we then have
\begin{equation}
\alpha_1 + \beta_1 \rightarrow 2B\,\sqrt{{3\over 2} - \nu}  \rightarrow 0\,.
\end{equation}
Therefore the apparently state dependent contribution (\ref{eq:epsone}) 
from the $k=1$ mode {\it vanishes} in any infrared finite state parameterized 
by finite $A$ and $B$, while the $k>1$ contribution is still given
by (\ref{eq:kgone}), and we have the result
\begin{equation}
\varepsilon_R =\lim_{\xi \rightarrow 0}\varepsilon_{BD}(\xi)\Big\vert_{m =0}
- {3\over 32\pi^2} 
\left({R\over 12}\right)^2= \varepsilon_{AF}
\; , 
\end{equation}
in any $(A,B)$ ``vacuum" state with $\alpha_k =1, \beta_k = n_k = 0$
for $k>1$, upon restoring physical units. The
difference here is just that required to give the Allen-Folacci (AF)
renormalized expectation value, provided the limit of the BD value is
taken in the same order of $m=0$, $\xi \rightarrow 0$ that we have
evaluated (\ref{eq:epsone}). Since from (\ref{eq:tmunu-bd})
\begin{equation}
\varepsilon_{BD}(\xi)\Big\vert_{m =0} = {3\over 16\pi^2} 
\left({R\over 12}\right)^2 
\left[{1\over 180} - {1\over 6}(6\xi -1)^2\right]\,,
\end{equation}
and both the BD and AF values are de Sitter invariant,
\begin{eqnarray}
\lim_{\xi \rightarrow 0}\langle T_{ab}\rangle_{ren}(\xi)\Big\vert_{m =0} &=& 
- {3 g_{ab}\over 16\pi^2}\left({R\over 12}\right)^2
\left( {1\over 180} - {1\over 6} - {1\over 2}\right)\nonumber\\ 
&=& \frac {119\ }{960\pi^2}  \left({R\over 12}\right)^2 g_{ab}
= \frac{119\ R^2}{138\, 240 \pi^2}g_{ab} = \langle T_{ab}\rangle_{AF}
\; ,
\end{eqnarray}
and we obtain the de Sitter invariant AF result
for all $(A,B)$ ``vacuum" states in the massless, minimally coupled case.

With the above careful analysis of the spatially homogeneous mode and
the independence of the asymptotic value of its energy-momentum tensor in an
arbitrary $k=1$ infrared finite state characterized by $A$ and $B$, it is
now straightforward to carry out the proof of the attractor behavior
of the AF state for an arbitrary UV finite physical state with
$m=\xi=0$, by allowing the $k>1$ modes to have $\beta_k$ and $n_k$
different from zero.  Substituting (\ref{eq:psik}), (\ref{eq:psiknon}), 
(\ref{eq:psionelim}), and (\ref{eq:psikone}) into (\ref{eq:T00u}), 
(\ref{eq:Tru}) with $m=\xi=0$, and using (\ref{eq:AkBk}) gives
\begin{mathletters}
\begin{eqnarray}
\varepsilon = -\langle {T^0}_0 \rangle_{ren} &=& 
-\langle {T^0}_0\rangle_{AF}
 +(1+ 2n_1) \frac{|A|^2 \cos^6 \eta}{\pi^2}
\nonumber \\
&  &+
\;  \frac{1}{4 \pi^2}
\sum_{k=2}^{\infty} \left\{ \left[
2 n_k + 2 (1+2n_k) |\beta_k|^2
\right]
[k^3 \cos^4 \eta
+ k ( -\cos^4 \eta + \frac{1}{2} \cos^2 \eta)] \right.
\nonumber \\
  &  & +
\;
(1+ 2n_k) \left[
 (\beta_k \alpha_k^* e^{2 i k \eta} + \beta_k^* \alpha_k e^{-2ik\eta})
   k ( -\cos^4 \eta + \frac{1}{2}  \cos^2 \eta)
\right.
\nonumber \\
  &  & \left. \left.
+ \;  i
\;  (\beta_k \alpha_k^* e^{2ik\eta} - \beta_k^* \alpha_k e^{-2ik\eta})
k^2 \cos^3 \eta
     \sin \eta \right]\right\}
\; ,
\label{eq:t0032}  \\
\langle T\rangle_{ren} &=& \langle T\rangle_{AF}
+ (1+2n_1) \frac{2 |A|^2 \cos^6 \eta}{\pi^2}
\nonumber
\\
&& - \; \frac{1}{4 \pi^2} \sum_{k=2}^{\infty}
\left\{
\left[ 2 n_k + 2 (1 + 2n_k) |\beta_k|^2
\right]
k
\cos^2 \eta
\right.   \nonumber \\
& &
+
\; (1+2n_k)
\left[(\beta_k \alpha_k^* e^{2ik\eta} + \beta_k^* \alpha_k e^{-2ik\eta}) 
(-2 k^3 \cos^4 \eta + k \cos^2 \eta)
\right.
\nonumber
\\
&& \left. \left.
+ \; 2 i \; (\beta_k \alpha_k^* e^{2ik\eta}
- \beta_k^* \alpha_k e^{-2ik\eta}) k^2 \cos^3 \eta
      \sin \eta \right]
\right\}
 \label{eq:tr32}
\; .
\end{eqnarray}
\end{mathletters}

Provided the $k$ sums converge, it is clear that all the state
dependent terms contain at least one factor of $a^{-2} = \cos^2\eta$,
and so vanish in the limit of $\eta \rightarrow {\pi \over
2}$. However, the requirement that the state be fourth order adiabatic
just guarantees this convergence, for the same reason as in the
previous analysis in spatially flat coordinates. Indeed we have
\begin{eqnarray} 
\vert\beta_k\vert &= & \frac {C(k)} {k^4}\,,\nonumber\\
n_k &=& \frac {N(k)} {k^4}\,,
\end{eqnarray}
for some $C(k)$ and $N(k)$ that vanish as $k\rightarrow \infty$.
This is sufficient to guarantee the absolute convergence of all terms 
in the sums. Since all state dependent terms
are multiplied by at least two powers of $\cos \eta = a^{-1}$, which
vanishes in the late time limit $\eta \rightarrow {\pi \over 2}$, we
conclude that any UV and IR admissible state of the massless,
minimally coupled scalar field has an energy-momentum tensor which
approaches the AF value, $\langle T_{ab} \rangle_{AF}$ in
the late time limit $\eta \rightarrow {\pi\over 2}$.

If one considers the contributions of the state dependent terms to the
energy density and trace from the $k>1$ modes, for $\nu$ not exactly
$3\over 2$, the kinematics is essentially the same as our previous
analysis of the $k=1$ mode.  Their contribution also falls off
proportional to $\xi |\alpha_k + \beta_k|^2 a^{2\nu - 3}$ at late
times for $\nu$ close to $3\over 2$. However, there is no compensating
large factor coming from the pole in the $\Gamma$ function
normalization constant as there is for the $k=1$ mode.  Hence, the
coefficient of this slow fall off goes to zero as the massless,
minimally coupled limit is approached. That is, exactly at $m=\xi=0$
when the contribution of the $k>1$ modes no longer falls off and they
could in principle contribute to the asymptotic value of $\langle
T_{ab}\rangle_{ren}$ at late times, at that very point their
coefficient vanishes identically and they make no contribution at
all. Thus, at precisely $m=0$ and $\xi =0$ the difference between the
BD and AF values can be attributed entirely to the additional
condensate in the spatially homogeneous $k=1$ mode alone, and there
are no slow transient modes in our explicit analysis of the massless,
minimally coupled energy and trace.

These considerations are relevant to the case when $\nu = {3\over 2}$
but $m$ and $\xi$ are separately non-zero. The calculation is almost
identical but the conclusion is different, since now the finite $k>1$
mode sum is multiplied by a coefficient $\xi$ which does {\it not}
vanish. The entire mode sum of state independent terms from $k = 2$ up
to $a(\eta)$ do not fall off at late times and in fact all add up, to
give a contribution proportional to $\langle \Phi^2\rangle
\sim\sum_{k=2}^a k^{-1} \sim \log a$ in $\langle T_{ab}\rangle_{ren}$,
which grows linearly in comoving time.  The explicit expression is
most conveniently calculated by using (\ref{eq:deftan}) to divide the
energy-momentum tensor into a state dependent numerical part, $\langle
T_{ab} \rangle_n$ and a state independent analytic part, $\langle
T_{ab} \rangle_{an}$.  They are separately conserved~\cite{AE}.  The
quantities $\langle T_{ab} \rangle_d $ and $\langle T_{ab}
\rangle_{an} $ are given in Ref.~\cite{AE}.  By substituting
Eqs. (\ref{eq:psikone}) and (\ref{eq:psiknon}) into
Eqs. (\ref{eq:T00u}) and (\ref{eq:Tru}), and subtracting the expression
for $\langle T_{ab} \rangle_d $ given in Ref.~\cite{AE}, we find that
$\langle T_{ab} \rangle_n$ approaches a state dependent, finite
constant in the limit $\eta \rightarrow {\pi\over 2}$. However, the
quantity $\langle T_{ab} \rangle_{an} $ has a term that is
proportional to $\langle \Phi^2\rangle$. At late times this term
behaves as $\log a$ and dominates. In fact
\begin{mathletters}
\begin{eqnarray}
\varepsilon_{ren}
&\rightarrow & -\frac{3 \xi}{4\pi^2} \log (\cos \eta)
+ q_1
\; ,     \label{eq:q1} \\
\langle T \rangle_{ren}  &\rightarrow & {3 \xi  \over {\pi^2}}
 \log (\cos \eta)
+ q_2
\;, \label{eq:q2}
\end{eqnarray}
\end{mathletters}%
with the constants $q_1$ and $q_2$ dependent on the state of the field
and constrained by the conservation equation.  Since $\log (\cos \eta)
\rightarrow t$ in comoving time, the energy-momentum tensor grows
linearly with $t$ at late times.  
For $\xi >0$ (and $m^2<0$) the linear growth in comoving time
decreases the effective cosmological ``constant", whereas for $\xi <0$
(and $m^2>0$) it increases it.  In either case the back-reaction of
the energy-momentum of the quantum scalar field for $m^2 + \xi R = 0$
(but $m^2 = -\xi R \neq 0$) on the geometry certainly cannot be
neglected at late times since $\langle T_{ab}\rangle$ grows without
bound for any physical state. The fact that a massive non-minimal
field with $\xi <0$ can induce an effective cosmological ``constant"
due to inflationary particle production was noted in Ref.~\cite{sh}
using a different approach.  It is an interesting open question
whether this linearly growing behavior (in proper time) carries over
to the physically more relevant case of one-loop gravitons, since the
mode functions for gravitons in a particular gauge obey the same
equation in a RW spacetime as do the mode functions for a massless
minimally coupled scalar field~\cite{Grischuk}.

Finally, we note that for all real values of $\nu$, the analysis of
the $k=1$ mode in this Section may be extended to all of the higher
$k$ modes, since the late time behavior of the higher $k$ modes is
determined by that of the hypergeometric function in (\ref{eq:hyper}),
which gives 
\begin{equation}
f_k \left(\eta \rightarrow {\pi\over 2}\right) \rightarrow 
\left[ {\Gamma \left( k+ {1\over 2} - \nu\right) \over 2 \Gamma 
\left( k + {1\over 2} + \nu\right)}\right]^{1\over 2} 
{\Gamma (2 \nu) \over \Gamma ({1\over 2} + \nu)}
{(-i)^k\over k!}\left({i\sec \eta\over 2}\right)^{\nu - {1\over 2}} \,.
\end{equation}
Thus all of the higher $k$ modes behave as $a^{\nu - {1\over 2}}$ and
give the (unrenormalized) energy density the leading order late time behavior
\begin{equation}
a^{2\nu - 3}\left[\left(\nu -{3\over 2}\right)^2 + 12\xi (\nu -1) + m^2\right]
\left[{\Gamma(2\nu)\over 2^{\nu}\Gamma\left(\nu + {1\over 2}\right)}\right]^2
\sum_{k = 1}^{\infty} (1 + 2n_k){\Gamma \left( k+ {1\over 2} - \nu\right) 
\over \Gamma \left( k + {1\over 2} + \nu\right)(k!)^2}\vert\alpha_k +
e^{i \pi (k + \nu -{1\over 2})}\beta_k\vert^2\,,
\label{eq:div}
\end{equation}
which grows (or shrinks) like  $a^{2\nu - 3}$ for an arbitrary physical state
unless either the coefficient vanishes or $\alpha_k + e^{i\pi \left(k
+ \nu - {1\over 2}\right)}\beta_k = 0$ for all $k$. However, the
latter is impossible since it is inconsistent with the requirement
that all physically allowable states be fourth order adiabatic states,
which requires that $\alpha_k \rightarrow 1$ and $\beta_k \rightarrow
0$ sufficiently fast as $k \rightarrow \infty$, in order for the
renormalized energy density to be UV finite. Since the fourth order
adiabatic subtraction behaves at most like a constant at late times,
the $a^{2\nu - 3}$ behavior is not affected by the UV subtraction, and
indeed the sum in (\ref{eq:div}) converges at large $k$ for any UV
admissable state. Hence after the adiabatic four subtraction the
leading order late time behavior indicated by (\ref{eq:div}) survives
unless the first bracket in front of the entire expression
vanishes. The quantity in this bracket is identical to the factor in
(\ref{eq:lateps}) found in the flat spatial section analysis of the
last Section. The corresponding quantity for the trace is given by the
factor in curly brackets in (\ref{eq:lateT}), which is a similar
combination of $\nu, \xi$, and $m^2$. If we require that both of these
factors vanish identically, to eliminate the leading order behavior in all
components of $\langle T_{ab}\rangle_{ren}$, then these two conditions
plus the defining relation (\ref{eq:nu}) give
\begin{eqnarray}
 m^2 &=& - \frac{\nu (2 \nu - 3) (2 \nu - 1)}{4 (\nu - 2)}\; , \nonumber \\
 \xi &=& \frac{(2 \nu - 3)}{8 (\nu - 2)}
\; .
\end{eqnarray}
Thus, except for $\nu = 2$, for any given $\nu$ there is always one
value of $m^2$ and one value of $\xi$ for which the coefficients of
these leading order terms in $\langle T_{ab}\rangle_{ren}$ vanish.  The next
to leading order terms go like $a^{2\nu -5}$, and grow without bound at late 
times in any case when $\nu > {5\over 2}$.  

The above analysis implies that, for most values of $m^2$ and $\xi$, when
$\nu > {3\over 2}$ the leading order terms in the
components of the energy-momentum tensor grow without bound like
$a^{2\nu - 3}$ in de Sitter space for any physically admissable
initial state of the scalar field.  These values of $\nu$ correspond
to the purely tachyonic cases $m^2 + \xi R <0$.  

\section{Numerical Studies}
\label{sec:numerical}

In this Section we display numerical results for various values of
$\nu$.  All results are given in dimensionless units where $\alpha =
1$ and $R = 12$.  The quantum state in each case is a fourth order
adiabatic state matched to the vacuum at some initial time
$\eta_0$. That is, we choose initial conditions for the mode function
and its first derivative to be
\begin{eqnarray}
\psi_k (\eta_0) &=& \psi_k^{(4)}(\eta_0)\,,\nonumber\\
{{\rm d} \psi_k\over {\rm d}\eta}(\eta_0) &=& {{\rm d} 
\psi_k^{(4)}\over {\rm d}\eta}(\eta_0)\,,
\end{eqnarray}
with $\psi_k^{(4)}$ the fourth order adiabatic mode function defined
by (\ref{eq:psiadb}) with phase measured from the initial time
$\eta_0$.  The initial time was chosen to be $t_0 =1$ in comoving
coordinates, {\it i.e.} $\sec \eta_0 = \cosh (1) = 1.54308\dots$.

For $\; \Re (\nu)<{3\over 2}$ the proof in Section \ref{sec:analytic} states
that the energy-momentum tensor approaches the Bunch-Davies value
(\ref{eq:tmunu-bd}) at late times for an arbitrary physically
admissable state.  This occurs due to a redshifting of the
state dependent part of the energy-momentum tensor. The initial
state dependent transient contributions fall off like
\begin{equation}
a^{2 \nu -3} = (\cosh t)^{2 \nu - 3} \;.
\end{equation}
Thus the characteristic time to approach the BD value is
\begin{equation}
\tau = \frac{1}{3 - 2 \nu}\,.
\end{equation}
In Fig. 1 we plot the renormalized energy 
density for these adiabatic initial conditions with $m=0$ and 
$\xi = {1\over 7}$. For this relatively large value of $\xi$, the 
characteristic time $\tau$ is of order one and the energy density 
approaches the BD value within one expansion time.

For smaller values of $\xi$ the initial value transients persist for
longer times. Our analysis in the previous Section shows that the
$k=1$ contribution is essential for the shift from the BD to
AF value as $\xi \rightarrow 0$. Since the $k=1$ mode in an arbitrary
physical state contributes to the energy density the value 
$\varepsilon_1$ given by (\ref{eq:epsone}), up to terms which fall 
off like $a^{-2}$, for arbitrary $\alpha_1, \beta_1$, and $n_1$, 
while the BD state has $\alpha_1 =1, \beta_1 = n_1 =0$, 
it is clear that the difference of the renormalized energy density
from the BD value coming from this mode is
\begin{equation}
\varepsilon_R - \varepsilon_{BD} = \varepsilon_1 - {3\over 32\pi^2} 
\left({R\over 12}\right)^2 a^{2 \nu - 3}\,.
\end{equation}
For the adiabatic initial conditions here we find that
\begin{equation}
\psi_1(\eta_0) = {1 \over \sqrt{2 W_1^{(4)}(\eta_0)}}
\approx {(\alpha_1 + \beta_1) \sec\eta_0\over
2 \sqrt{{3\over 2} - \nu}}\,.
\end{equation}
Since $W_1^{(4)}(\eta_0)$ remains finite as $\xi \rightarrow 0$,
provided that $\eta_0 \neq 0$, it follows that in this limit
$|\alpha_1 + \beta_1|^2 \sim 3 - 2 \nu \sim \xi$ also goes to zero. 
Hence for small $\xi$, $\varepsilon_R$ goes to a value close to (
but slightly larger than) the AF value after a time of order one. 
This is observed in both Figs. 2 and 3 for $\xi = {1\over 100}$ and 
$\xi = {1\over 1000}$, respectively.
\begin{figure}
\epsfxsize=15cm
\epsfysize=8cm
\centerline{\epsfbox{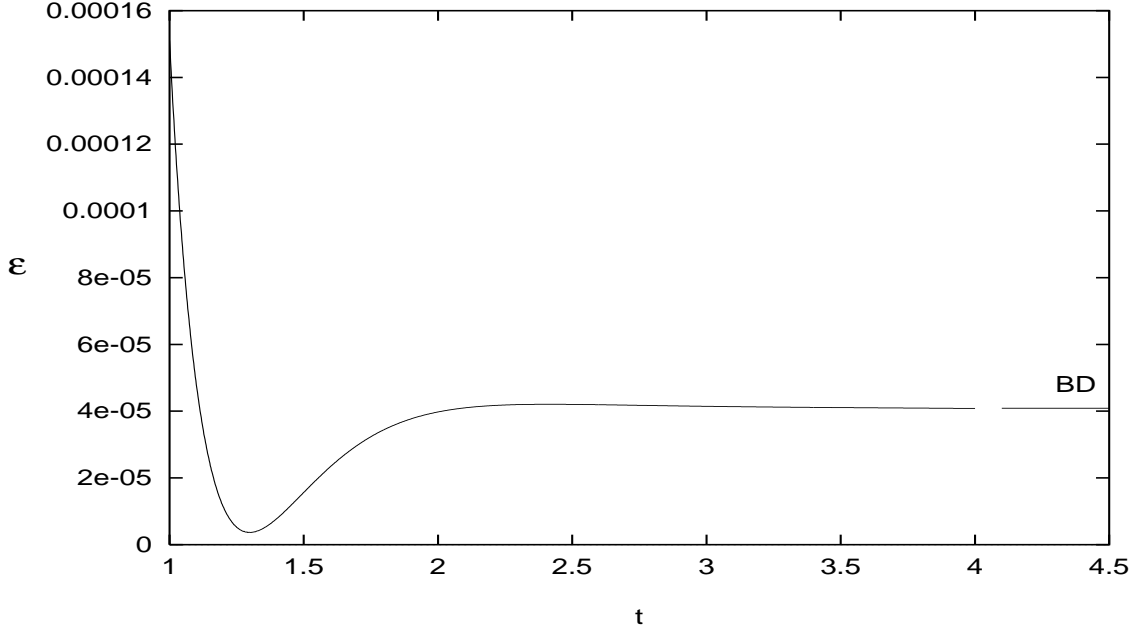}}
\vspace{0.35cm}
\caption
{The energy density for $m=0$ and $\xi={1\over 7}$ as a function of
comoving time $t$ for adiabatic initial conditions at $t=1$.  The
attractor behavior of the Bunch-Davies value for the energy density is
illustrated.}
\label{graph7}
\vspace{0.2cm}
\end{figure}
\begin{figure}
\epsfxsize=15cm
\epsfysize=8cm
\centerline{\epsfbox{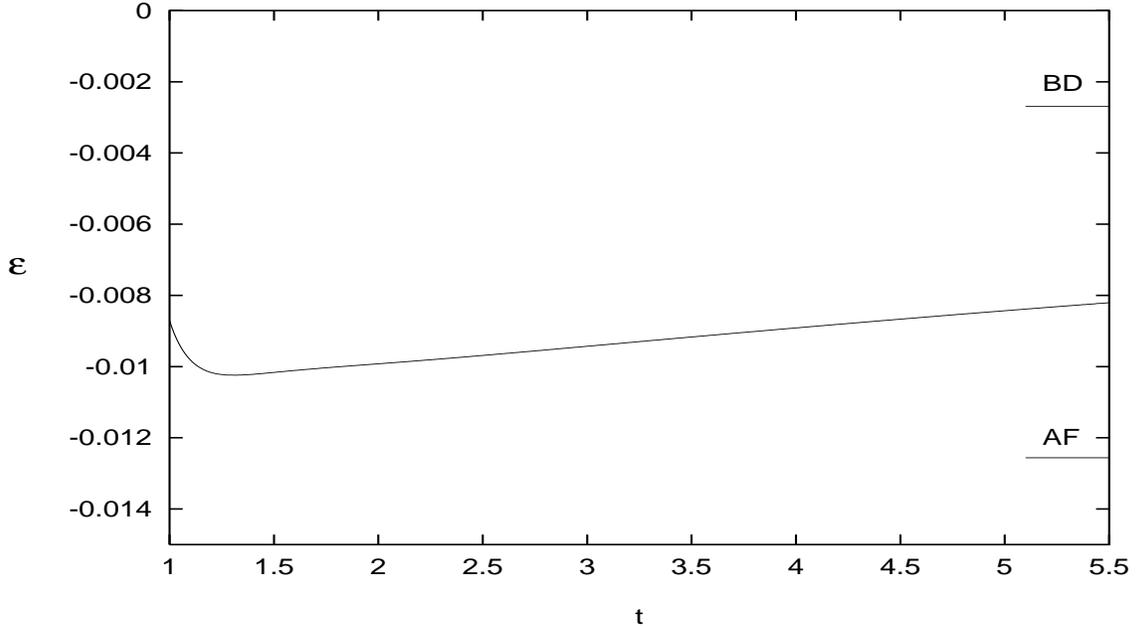}}
\vspace{0.35cm}
\caption
{The energy density for $m=0$ and $\xi={1\over 100}$ as a function 
of comoving time $t$ for adiabatic initial conditions at $t=1$. 
The line segments at right giving the values of
the energy density for the Allen-Folacci (AF) state,
and the Bunch-Davies (BD) state (for $m=0$ and $\xi={1\over 100}$).}
\label{graph100}
\end{figure}
\begin{figure}
\epsfxsize=15cm
\epsfysize=8cm
\centerline{\epsfbox{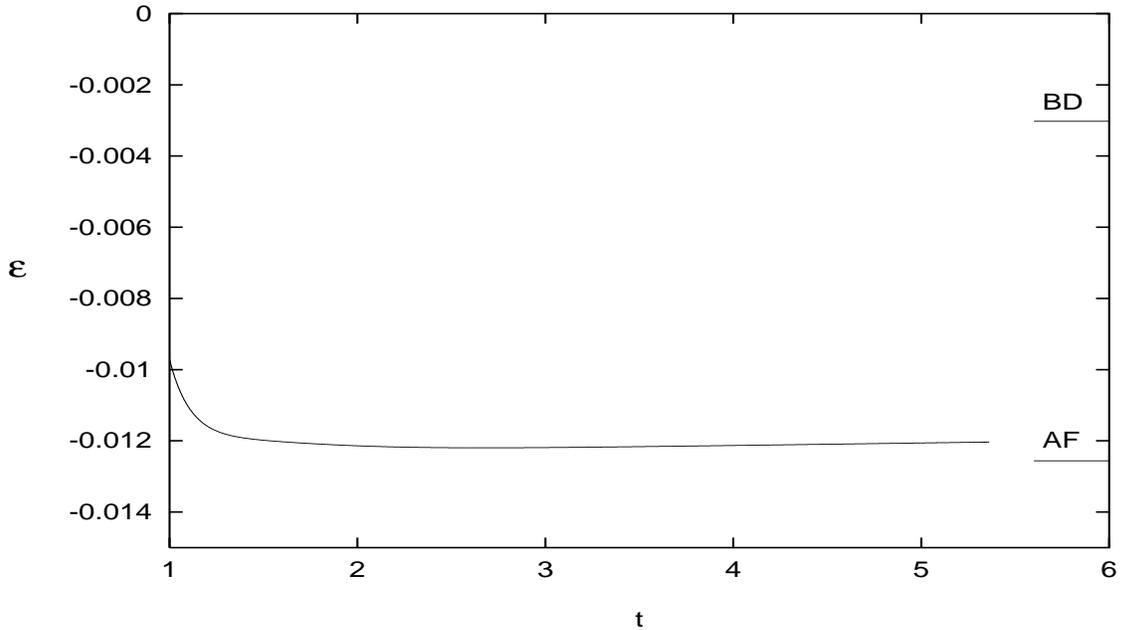}}
\vspace{0.35cm}
\caption
{The energy density for $m=0$ and $\xi={1\over 1000}$ for the same
initial conditions as the previous figures. Notice the very
long-lived transient behavior of $\varepsilon$, staying close to
the AF value with a small positive linear slope towards the BD value.}
\label{graph1000}
\vspace{0.2cm}
\end{figure}

Further, the energy-momentum tensor contains a term proportional to
$\xi \langle \Phi^2\rangle$ which would grow linearly in comoving time
for $\nu$ very close to $3\over 2$, except for the factor of $a^{2 \nu
- 3}$ that damps it to zero at very late times $t \gg \tau$.  Hence on
times $1 < t \ll \tau$, where the $a^{2 \nu - 3}$ factor is
essentially constant, we should expect this linear growth of
$\varepsilon$ in comoving time with a slope proportional to $\xi$,
given by (\ref{eq:q1}). This behavior is demonstrated in Figs. 2 and
3.  In the latter case $\xi$ is so small that $\alpha_1 + \beta_1$ and
the slope are nearly vanishing and the energy density stays close to
the AF value until times of order $\tau = 125$, which is much larger
than the times shown in Fig. 3. When $\xi =0$ (still keeping $m=0$)
both $\alpha_1 + \beta_1$ and the slope vanishes identically, so the
energy-momentum tensor goes to the AF value and remains there.  This
demonstrates that the limit $\Re (\nu)\rightarrow {3\over 2}$ is quite
continuous when viewed at finite times with well-defined physical
initial conditions, although the late time limit is discontinuous.

\begin{figure}
\epsfxsize=15cm
\epsfysize=8cm
\centerline{\epsfbox{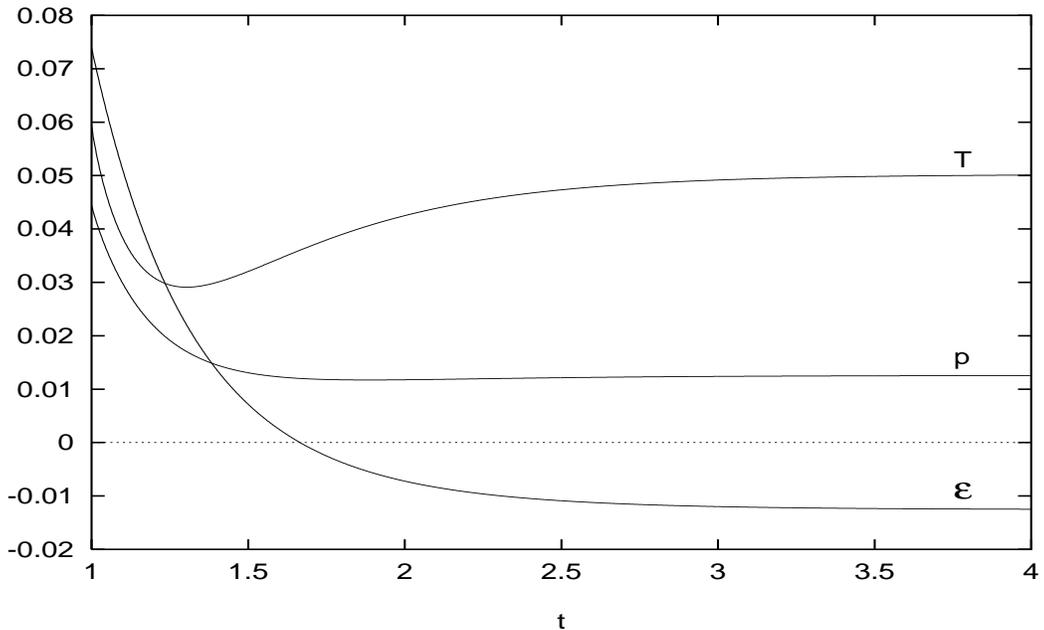}}
\vspace{0.35cm}
\caption
{From top to bottom the curves are the trace, pressure, and energy
density for $m=0$ and $\xi=0$. The state is a fourth order adiabatic
state at $t=1$ with $n_1 = n_2 = 2$ and $n_k = 0$ for all $k > 2$. The trace,
pressure, and energy density approach their AF fixed point
values at late times.}
\label{sigma2}
\end{figure}

In the case $\nu = {3\over 2}$ there are two distinct behaviors
depending on whether $m$ and $\xi$ are separately vanishing or not.
In the massless, minimally coupled case we proved in Section \ref{sec:nu32}
that the Allen-Folacci value is a fixed point at late times.
In Figure 4 we show the approach of the
energy density, pressure, and trace to their Allen-Folacci values for
the massless minimally coupled field.  The field is in an
``n-particle'' state with $n_1 = n_2 = 2$ and $n_k = 0$ for all $k > 2$.
In Figure 5 we show the behavior of the energy density and trace for
the case $m = 0.3$ and $\xi = - 0.0075$ when the field is in a fourth
order adiabatic vacuum state.
\begin{figure}
\epsfxsize=15cm
\epsfysize=8cm
\centerline{\epsfbox{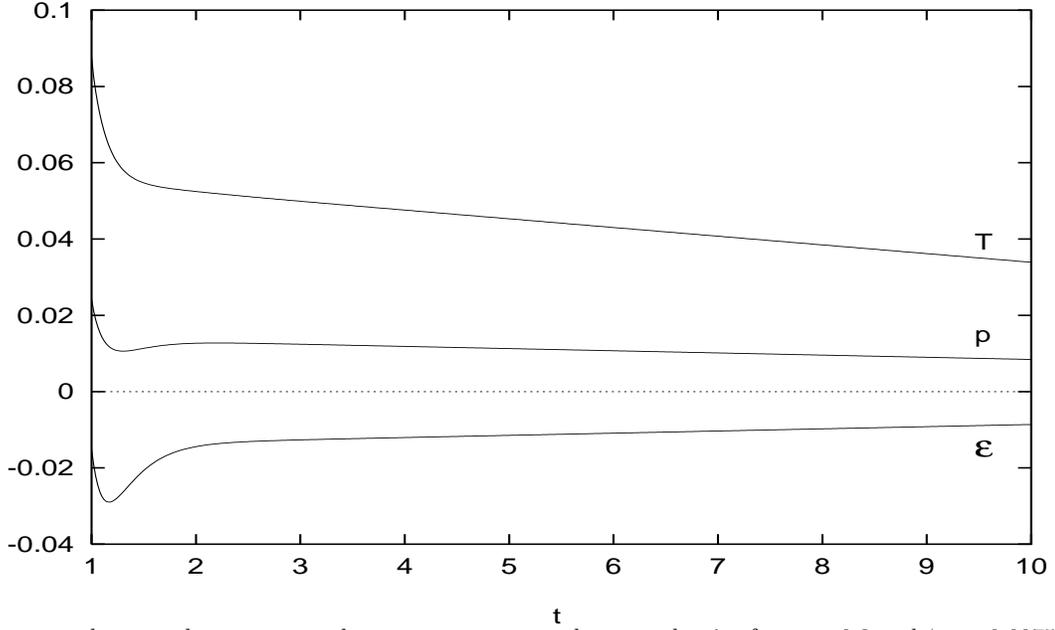}}
\vspace{0.35cm}
\caption
{From top to bottom the curves are the trace, pressure, and energy
density for $m=0.3$ and $\xi=-0.0075$, $n_k =0$ and the fourth
order adiabatic state at the initial time $t=1$.  The curves behave linearly
in comoving time $t$ at late times, with their slopes given by 
Eqs. (\ref{eq:q1}) and (\ref{eq:q2}).}
\label{nu32}
\vspace{0.2cm}
\end{figure}
\begin{figure}
\epsfxsize=15cm
\epsfysize=8cm
\centerline{\epsfbox{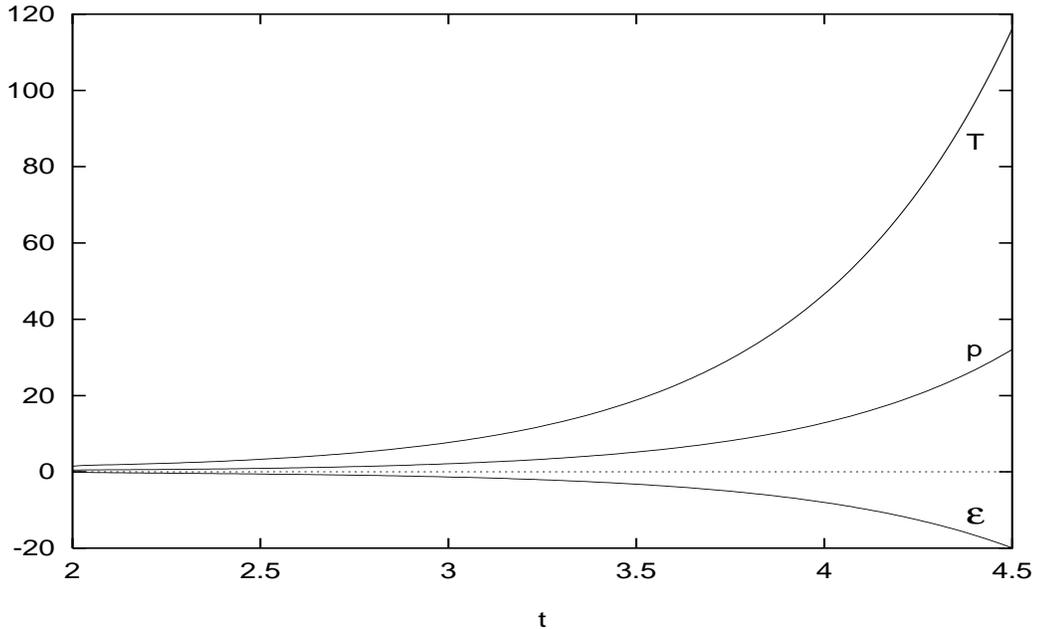}}
\vspace{0.35cm}
\caption
{From top to bottom the curves are the trace, pressure, and energy
density for $m=0.1$ and $\xi=-0.3$, $n_k =0$ and the fourth
order adiabatic state at the initial time $t=1$.}
\label{tensorb-3}
\end{figure}

When $\Re (\nu) > {3\over 2}$ the analysis at the end of Section
\ref{sec:nu32} shows that the leading order state dependent terms in
the energy-momentum tensor will generally grow exponentially with
time. This is illustrated in Fig. 6. The exponential growth that
occurs here is similar to the well known one for the classical scalar
field when $m$ and $\xi$ have values such that $\nu>{3\over
2}$~\cite{dolgov,ford}. Such fields are tachyonic and presumbably of
little physical interest unless interactions are added to stabilize
them.

\section{Infrared Scaling and the Generalized Anomaly}
\label{sec:scaling}

We have found that for all $\Re (\nu)<{3\over 2}$ the renormalized
expectation value of $\langle T_{ab} \rangle$ approaches the de Sitter
invariant Bunch-Davies value for any physically admissable initial
state, whereas it approaches the de Sitter invariant Allen-Folacci
value for any physically admissable initial state in the massless,
minimally coupled case. Since all the initial state dependence
vanishes asymptotically, these state independent de Sitter invariant
fixed point values for $\langle T_{ab} \rangle_{ren}$ must be purely
geometrical in origin. Indeed, both the Bunch-Davies point-splitting
calculation and the Hadamard calculation of Allen-Folacci rely only on
the properties of the two-point function of the scalar field $G(x,x')$
for $x' \rightarrow x$. Hence, the BD and AF asymptotic values of
$\langle T_{ab} \rangle_{ren}$ are certainly ``pseudo-local'' in the
terminology of Ref.~\cite{dfcb}, {\it i.e.} they are expressible in
terms of purely {\it local} functions of the RW scale factor $a(\eta)$
and its derivatives.

If we specialize now to zero mass, $m=0$, then on simple dimensional
grounds the asymptotic $\langle T_{ab} \rangle_{ren}$ can be expressed
purely in terms of local conserved tensors of fourth adiabatic order.
Although we have used adiabatic subtraction methods to renormalize
$\langle T_{ab} \rangle$ it is known that the value of $\langle T_{ab}
\rangle_{ren}$ so obtained is equal to that in a fully covariant
procedure such as dimensional regularization or covariant
point-splitting~\cite{AP,Birr}.  In a fully covariant procedure, which
yields a local conserved tensor of fourth adiabatic order, only the
Riemann tensor together with its covariant derivatives and
contractions can appear. Hence $\langle T_{ab} \rangle_{ren}$ for
$m=0$ must be expressible entirely in terms of such local geometrical
tensors.

In four dimensions the only such local tensors are linear
combinations of $^{(1)}H_{ab}$, $^{(2)}H_{ab}$, and $^{(3)}H_{ab}$, where 
\begin{eqnarray}
^{(1)}H_{ab} &\equiv& {1\over \sqrt{-g}} {\delta\over\delta g^{ab}}
\int\, \sqrt{-g}\,R^2\,{\rm d}^4x 
\nonumber \\
&=& 2 g_{ab} \sq R - 2\nabla_a\nabla_b R  
 + 2 R R_{ab} - {1\over 2} g_{ab} R^2\,,
\label{eq:Hone}
\end{eqnarray}
vanishes in de Sitter spacetime and
\begin{equation}
^{(3)}H_{ab} = R_a^cR_{cb} - {2\over 3}RR_{ab} -{1\over 2} R_{cd}R^{cd}g_{ab}
 + {1\over 4} R^2 g_{ab}\,.
\label{eq:Hthr}
\end{equation}
In RW spacetimes, which are all conformally flat, the tensor
$^{(2)}H_{ab}$ is proportional to $^{(1)}H_{ab}$ and hence vanishes as
well.  Therefore the only non-trivial fourth order conserved
geometrical tensor in de Sitter spacetime is $^{(3)} H_{ab}$ and we
conclude that the fixed point BD and AF values found in our previous
analysis are proportional to
\begin{equation}
^{(3)}H_{ab} = {R^2\over 48}\ g_{ab} = 3 \left({R\over 12}\right)^2 g_{ab}\,.
\label{eq:Hthrds}
\end{equation}
Furthermore, since
\begin{equation}
^{(3)} H_{ab}\,g^{ab} = - R_{ab}R^{ab} + {R^2\over 3}
={1\over 2} \left( R_{abcd}R^{abcd} - 4 R_{ab}R^{ab} + R^2\right)
 \equiv {G\over 2} \; , \end{equation} 
in RW spacetimes (where the Weyl tensor vanishes), the coefficient of
$^{(3)} H_{ab}$ is proportional to the coefficient of the Gauss-Bonnet
integrand in the trace of $\langle T_{ab} \rangle$. Such a term in the
trace is known to correspond to a non-local but nevertheless fully
covariant action and this action is precisely the same as that
generated by the trace anomaly of free conformal fields~\cite{amm}.
Since we have obtained fixed point results for the asymptotic values
of $\langle T_{ab} \rangle_{ren}$ for massless fields in de Sitter
space which are purely geometrical and proportional to $^{(3)}H_{ab}$,
even for {\it non}-conformal massless fields, we can give a definite
meaning to the value of the proportionality coefficient and the
non-local anomaly-like term in the effective action even when $\xi
\neq {1\over 6}$.

Let us define the generalized anomaly coefficient by fixing the
normalization
\begin{equation}
\lim_{t \rightarrow \infty} \langle T_{ab} \rangle_{ren} 
= -{Q^2\over 16\pi^2}\, 
^{(3)}H_{ab} = -{3 Q^2\over 16\pi^2}\left({R\over 12}\right)^2 g_{ab}\,.
\label{conom}
\end{equation}
With this normalization we find from the asymptotic value of $\langle
T_{ab} \rangle_{ren}$ for a scalar field in de Sitter space that
\begin{equation}
Q^2 = \left\{ \begin{array}{ll}
Q^2_{BD} &= {1\over 180} - {1\over 6} (6\xi -1)^2, \qquad m=0, \qquad \xi > 0\,,
\\
Q^2_{AF} &= {1\over 180} - {1\over 6} - {1\over 2} = -{119\over 180},
\qquad m=0, \qquad \xi = 0\,.
\end{array} \right.
\label{Qval}
\end{equation}
The value of $Q^2$ for a conformally invariant field
($m=0,\xi={1\over6}$) is $1\over 180$ and corresponds to the pure
trace anomaly coefficient. The first member of (\ref{Qval}) provides
the generalization of this coefficient away from the conformal case.
The discontinuous behavior at $\xi=0$ has been discussed in
Ref.~\cite{k-g}. As we have seen it arises from the singular behavior of the
spatially constant zero mode of the massless, minimally coupled field,
which is non-oscillatory and hence cannot be quantized as a Fock mode
in the same fashion as the oscillatory modes. We discussed 
this discontinuity in detail in Section \ref{sec:nu32}.

The connection of the tensor $^{(3)}H_{ab}$ with the trace anomaly
may be seen from the general form of the effective action for the
anomaly in a conformally flat space with metric~\cite{amm}
\begin{equation}
g_{ab} = e^{2\sigma} \bar g_{ab}\,,
\label{conf}
\end{equation}
namely
\begin{equation}
S_{\mathrm eff} =  -{Q^2\over 16\pi^2} \int\,
{\mathrm d}^4x \sqrt{-\bar g}
\left[\sigma \bar\Delta_4 \sigma + {1\over 2} \left(\overline G- {2\over 3}
{\,\raise.5pt\hbox{$\overline{\mbox{.09}{.09}}$}\,} 
{\overline R}\right)\sigma\right]\,,
\label{eq:effact}
\end{equation}
where $\Delta_4 = \sq ^2 + 2R^{ab} \nabla_a \nabla_b - {\textstyle
\frac{2}{3}} R {\sq}^2 + {\textstyle \frac{1}{3}} (\nabla^a R)
\nabla_a$ is the unique fourth order differential operator acting on
scalars which is conformally covariant. A fully covariant but
non-local form of the effective action (\ref{eq:effact}) can be
obtained by solving $\sqrt{-g}\left(G - {2\over 3} \sq R\right) =
\sqrt{-\bar g} \left(\bar G - {2\over 3} \sqb \bar R\right) +4
\sqrt{-\bar g} \bar \Delta_4 \sigma$ for $\sigma$. In that non-local
form all reference to the separation of the metric into background and
conformal factor as in (\ref{conf}) disappears.

The energy-momentum tensor following from the variation of the local
form of the effective action (\ref{eq:effact}) with respect to the
background metric $\bar g_{ab}$, is given by Eq. (2.9) of
Ref.~\cite{amm}. The form of this energy-momentum tensor simplifies
considerably on the Einstein space $R\times S^3$, where $\overline G =
{\,\raise.5pt\hbox{$\overline{\mbox{.09}{.09}}$}\,} {\overline R} =
0$.  If we set $\sigma=\log a(\eta)$ then this is equivalent to
evaluating the components of this energy-momentum tensor in a general
RW spacetime with closed spatial sections ($\kappa=+1$). Using the
expressions for $^{(1)}H_{ab}$ and $^{(3)}H_{ab}$ in a general RW
space in terms of $a(\eta)$ and its derivatives, we quickly find that
\begin{equation}
T_{ab}[\sigma] = -{2\over \sqrt{-g}} {\delta S_{\mathrm eff}\over 
\delta \bar g^{ab}}= {Q^2\over 16\pi^2}\left[ {1\over 18}\, ^{(1)}H_{ab} - 
^{(3)}H_{ab} + ^{(3)}H_{ab}\Big\vert_{R\times S^3}\right]\,.
\label{varT}
\end{equation}
Hence the tensor $^{(3)}H_{ab}$ which is called ``accidentally
conserved" in Ref.\ \cite{bi-da} is associated in fact with the
existence of a non-local covariant effective action related to the
trace anomaly, which has the local form (\ref{eq:effact}) when the
metric is conformally decomposed as in (\ref{conf}).

The last term in (\ref{varT}) would not have been present had we
varied the fully covariant but non-local form of the anomalous
effective action, in which all reference to the background $\bar
g_{ab}$ drops out. Equivalently, it is just canceled if we add to
(\ref{varT}) the Casimir energy on $R \times S^3$, which is determined
by the anomaly by a further conformal transformation from flat
space~\cite{brown}. In either case, we should drop this last term in
(\ref{varT}) which depends on the arbitrary background $\bar g_{ab}$.
Evaluating the first two terms on de Sitter space we find that it is
exactly the value of the asymptotic form of the renormalized
energy-momentum tensor $\langle T_{ab} \rangle$ in de Sitter space
(\ref{conom}), found previously, with the value of $Q^2$ for the
massless scalar given by (\ref{Qval}). Thus, the effective action
(\ref{eq:effact}), associated with the conformal trace anomaly,
appears in the effective action for a massless scalar field, even for
{\it non}-conformal couplings, $\xi\neq {1\over 6}$.

Since it is associated with the anomaly, the physical significance of
$S_{\mathrm eff}$ in the quantum effective action for the scalar field
is that it determines the scaling behavior of the field theory under
global Weyl transformations of the background space. Using the fact
that the Euclidean effective action is given by $I_{\mathrm
eff}=-S_{\mathrm eff}$ and that the Euclidean continuation of de
Sitter space is $S^4$ with Euler number $\chi=2$, we can vary
$I_{\mathrm eff}$ with respect to a constant $\sigma =\sigma_0$ and
obtain
\begin{equation}
{\partial I_{\mathrm eff}\over \partial \sigma_0} = \alpha 
{{\rm d} I_{\rm eff}\over {\rm d}\alpha} 
= {Q^2\over 32 \pi^2}\int_{S^4}\, {\mathrm d}^4x\,
\sqrt{g} \; G = Q^2 \chi = 2Q^2
\; ,
\label{eq:scalI}
\end{equation}
since ${{\rm d} \over {\rm d}\sigma_0} = {{\rm d}\over {\rm d}
\log\alpha}$ is a global rescaling of the $S^4$. As a check, this
relation can be verified explicitly by the $\zeta$ function evaluation
of the Euclidean effective action,
\begin{equation}
I_{\mathrm eff}= {1\over 2}\; {\rm Tr \, \log}
\left(-{\,\raise.5pt\hbox{$\mbox{.09}{.09}$}\,} + m^2 + \xi R\right) =
-{1\over 2}{{\mathrm d}\zeta\over {\mathrm d}s}
\Big\vert_{_{s=0}}\,,
\label{eq:Euceff}
\end{equation}
where the generalized zeta function for the Euclidean continuation of
the wave operator appearing in the Tr $\log$ is
\begin{equation}
\zeta (s) = \sum_{n=0}^{\infty} d_n \lambda_n^{-s}\; ,
\label{zdef}
\end{equation}
in terms of its eigenvalues $\lambda_n$ with degeneracy $d_n$ on
$S^4$. This sum is convergent for $\Re (s)>2$ and defines a meromorphic
function of $s$ which is analytic near $s=0$, where its derivative is
required. Introducing a mass scale $\mu$ to keep $\zeta (s)$
dimensionless for all $s$ leads in a standard calculation
to~\cite{mazmot}
\begin{equation}
-{1\over 2}{{\mathrm d}\zeta\over {\mathrm d}s}
\Big\vert_{_{s=0}} = -\zeta (0)
\log (\mu\alpha) + I_1(\nu)\,,
\label{Ieff}
\end{equation}
where $I_1(\nu)$ is a certain finite function of $\nu$, which
from the definition of $\nu$ (\ref{eq:nu}) becomes independent of
$\alpha$ when $m=0$, and the value $\zeta (0)$ is given by~\cite{mazmot}
\begin{equation}
-\zeta (0) = {1\over 12}\left( -\nu^4 + {1\over 2}\nu^2
+ {17\over 240}\right) = {1\over 90} - {1\over 3}(6\xi -1)^2
\; ,
\end{equation}
when $m=0$. By making use of (\ref{Qval}), (\ref{eq:Euceff}), 
and (\ref{Ieff}), we find that
\begin{equation}
\alpha{{\mathrm d}\over {\mathrm d}\alpha}
I_{\mathrm eff}(m=0) = -\zeta (0) = 2Q^2
\; ,
\end{equation}
is the behavior of the effective action for a massless scalar field
with $\xi>0$ under global Weyl rescaling of the metric, exactly as
predicted by(\ref{eq:scalI}) and the previous discussion based on 
the anomalous action (\ref{eq:effact}).

When $m=0$ and $\xi \rightarrow 0$, the integral representation of the
function $I_1(\nu)$ develops a logarithmic singularity, which can be
traced to the vanishing of the $n=0$ eigenvalue in the expression
(\ref{zdef}) for $\zeta (s)$. In this case the $n=0$ mode must be
excluded from the ultraviolet regulated sum over modes, which has the
effect of {\it adding} one unit to $\zeta (0)$ in the infrared scaling
behavior of the effective action~\cite{ammnp}, and accounts for the
addition of $-{1\over 2}$ in $Q^2$ in the minimally coupled
case. Hence the discontinuous behavior of $Q^2$ found in (\ref{Qval})
by our analysis of the asymptotic attractor behavior of the energy-momentum
tensor in de Sitter space is precisely the same as that
occuring in the effective action under global Weyl rescalings
when $m=0$ and $\xi \rightarrow 0$.

Note also that the global Weyl variation is given in terms of the
trace of the energy-momentum tensor by
\begin{equation}
\alpha{{\mathrm d}\over {\mathrm d}\alpha}
I_{\mathrm eff}(m=0) = -\int_{S^4}\, {\mathrm d}^4x\,\sqrt{g} \;
\langle T  \rangle_{ren} = -{8\pi^2\over 3}\alpha^4 
\langle T \rangle_{ren}\,.
\end{equation}
Since the $\zeta$ function method is fully covariant, the renormalized trace
$\langle T \rangle_{ren}$ computed in this way cannot contain non-covariant
contributions, and must be expressible entirely in terms of local
curvature invariants.

Finally we may consider relaxing the condition $m=0$.  If $m^2>0$, the
asymptotic form of the energy-momentum tensor for an arbitrary initial
state is given by the BD value. However, if we expand the BD result 
(\ref{eq:tmunu-bd}) in powers of $R/m^2$, we find that it contains 
{\it no} adiabatic order four $R^2$ terms, beginning instead with
$R^3/m^2$. Mathematically, this is because all terms up to fourth
adiabatic order have been removed by the ultraviolet regulating
procedure of point splitting or adiabatic subtraction. Hence the
coefficient of $^{(3)}H_{ab}$ at asymptotically late times in an 
arbitrary physical state is given by
\begin{equation}
Q^2 = 0 \; \; \; {\rm for} \; \; \;  m^2 > 0
\; ,
\label{eq:Qmzero}
\end{equation}
and no anomalous $S_{\mathrm eff}$ term appears in the quantum
effective action for a massive field. This is consistent with
the interpretation of the anomalous term (\ref{eq:effact}) in the 
effective action of the scalar field as an {\it infrared} effect, 
since the fluctuations of a massive field decouple 
at large distances or late times, and should induce only strictly 
irrelevant operators in the effective action in the far infrared, 
which are suppressed by positive powers of $R/m^2$. 

Only in the strictly conformal case, $m=0$ and $\xi = {1\over 6}$ is
the infrared effective action equal to that obtained by ultraviolet
methods, such as the $a_2$ coefficient in the Schwinger-DeWitt proper
time expansion. However, our analysis of the fixed point behavior of
$\langle T_{ab}\rangle_{ren}$ in de Sitter space shows that its
coefficient is connected with the global or extreme infrared scaling 
of the effective action, and this asymptotic behavior does not depend 
on the field being conformally invariant. The asymptotic attractor behavior 
of the energy-momentum tensor in de Sitter space defines an infrared scaling
coefficient that reduces to the trace anomaly coefficient in the
conformal case, but is a much more general concept than the trace
anomaly coefficient, since it is well-defined for all massless fields,
conformal or not. It is well-defined even for massive fields, although
as (\ref{eq:Qmzero}) shows, it vanishes in this case.

\section{Discussion and Conclusions}
\label{sec:discussion}

In this paper we have considered a quantum scalar field in a fixed de
Sitter background. We have studied the late time behavior of the
renormalized energy-momentum tensor and have found two important cases
in which, for arbitrary physically admissable states, the
energy-momentum tensor approaches a particular value at late times.
The values approached are those of the energy-momentum tensor in the
Bunch-Davies and Allen-Folacci states.  Thus, in this sense, these
special quantum states behave as fixed point attractors.

In the case $\Re (\nu) < {3\over 2}$ we have shown that for all fourth
order adiabatic states that are infrared finite the energy-momentum
tensor approaches the BD value at late times. The longest time scale
for the state dependent terms to redshift away is $\tau=(3 -2
\nu)^{-1}$.  This has been numerically verified for various values of
$m$ and $\xi$ when the fields are in a fourth order adiabatic state.
For the case in which $0 < {3\over 2} - \nu << 1$ and $m$ and $\xi$
are small, we numerically observe a more complicated behavior.  The
energy-momentum tensor quickly approaches the AF value and grows then
linearly with comoving time.  Our analytic proof implies that it must
then slowly decay to the BD value. It is worth noting that for $\Re
(\nu) > 0$ the redshift of the state dependent terms in the quantum
expectation value of $T_{ab}$ is slower than one might guess from the
redshifting of classical matter or radiation, {\it i.e.} $a^{-3}$ or
$a^{-4}$, respectively.

For the case $\nu={3\over 2}$ we have to distinguish two different
possibilities.  If the field is massless and minimally coupled then
we have proven that, for all fourth order adiabatic states that are
infrared finite, the energy-momentum tensor approaches the AF value at 
late times.  This is true for both vacuum and initially populated
states.  For any other values of $m$ and $\xi$
the renormalized energy-momentum tensor grows linearly with comoving
time, indicating that back-reaction effects need to be taken into
account. The sign of the linear growth depends on the sign of $\xi$.

For the tachyonic cases $\nu>{3\over 2}$ there is no attractor state.
Instead, for most values of $m^2$ and $\xi$ the renormalized
energy-momentum tensor grows like $a^{2\nu-3}$ at late times for all
physically admissable states.  Thus back-reaction effects again need
to be taken into account.

For the cases in which either the Bunch-Davies or Allen-Folacci states
serve as attractors in the above sense and the mass, $m$ of the field
is zero, we have shown how these results are connected to the
appearance of a certain non-local term in the quantum effective action
for the scalar field. This term gives rise to the local geometrical
tensor $^{(3)}H_{ab}$ in the asymptotic form of $\langle T_{ab}
\rangle$ at late times, and also determines the global scaling
behavior of the effective action for massless fields. Determining this
scaling behavior and relating it to the asymptotic $\langle T_{ab}
\rangle$ in de Sitter space has allowed us to generalize the notion of
the trace anomaly to massless, non-conformally coupled scalar fields,
in the sense that the coefficient of this non-local term in the
effective action is well defined even for $\xi \neq {1\over 6}$.

The interplay between the UV and IR properties of the state and the
mode sums contributing to the energy-momentum tensor is a theme
running through all of these results. We had to be careful to remove
the UV divergences from the unrenormalized $\langle T_{ab} \rangle$ in
order to analyze the late time limit. Since we have found state
independent de Sitter invariant results in both the $\nu < {3\over 2}$
and massless, minimally coupled cases, and since all the state
dependence resides in the finite $k$ modes, their form at high $k$
being restricted by the requirement of matching the adiabatic order
four vacuum, it is clear that state independent results for $\langle
T_{ab}\rangle_{ren}$ are possible only because the contribution to the
BD or AF expectation value comes from arbitrarily large $k$ at very
late times. In fact, inspection of the renormalized expectation value
of $\langle T_{ab}\rangle$ expressed as a mode sum, after the fourth
order adiabatic subtraction has been made, shows that the finite
contribution comes from $k \sim a$, {\it i.e.} when the physical
wavelength of the mode is of order of the de Sitter horizon.  At very
late times, this corresponds to arbitrarily large values of the
coordinate wave number $k$.

The finite difference between the BD value and the AF value in the
$m=\xi =0$ case comes entirely from the $k=1$ mode in closed spatial
sections, which is a purely IR effect. This leads to a finite
discontinuity in the infrared scaling properties of a massless field
since the value of the energy-momentum tensor at $m = \xi = 0$ is
different from its value in the limit $m=0$ and $\xi \rightarrow 0$.
The appearance of the $^{(3)}H_{ab}$ tensor and the corresponding
non-local action is quite generic for massless fields; its coefficient
$Q^2$ vanishes only if the mass is non-zero. Hence we should expect
that, although they are certainly not conformal, gravitons will also
contribute to this same infrared effective action with a finite value
of $Q^2$, which can be determined in the same way by a background de
Sitter calculation of their quantum $\langle T_{ab} \rangle$ at late
times. We plan to present the results of this calculation in a future
publication.

It is also interesting to note that the coefficient of the generalized
trace anomaly, $Q^2$, is not generically positive, in contrast to all
the previously known examples of massless conformal
fields~\cite{Duff}. It appears that the reason for this is that a
positive $Q^2$ comes from the ultraviolet behavior of $\langle T_{ab}
\rangle$ for conformal fields, while the infrared behavior of
non-conformal fields can contribute a negative value.  For any $Q^2
\neq 0$ the new term in the effective action leads to dynamics for the
RW scale factor which is quite different from the Einstein theory, and
remains to be investigated in a full dynamical back-reaction
calculation.

\acknowledgements 

Several of us would like to thank V. Sahni for helpful discussions,
and the Institute for Nuclear Theory, University of Washington, where
some of this work was completed. P. R. A. would like to thank T-8, Los
Alamos National Laboratory for its hospitality.  This work was
supported in part by grant numbers DMR-9403009 and PHY-9800971 from
the National Science Foundation.  It was also supported in part by
contract number W-7405-ENG-36 from the Department of Energy.

\appendix

\section{The cases of imaginary and integer values of $\nu$}

In this appendix we show that the proof in Section III works for
imaginary values of $\nu$ as well as for $\nu = 0,1$.  For imaginary
values of $\nu$ it is useful to write $\nu = i \gamma$ with $\gamma$ a
positive real number.  Then the formulas in Eq.\ (\ref{eq:I00}) and
(\ref{eq:I}) are the same as before.  However the values of the $A_i$ are
different.  They are given by
\begin{eqnarray}
A_1 &=& \frac{\pi}{2 k} e^{-\gamma \pi} \left[{\rm coth}(\gamma\pi) 
{\rm csch}(\gamma\pi) ((1+2 n_k)|c_2|^2 + n_k) - \frac{1}{2} 
{\rm coth}(\gamma\pi)
{\rm csch}(\gamma\pi) (1+ 2n_k) (c_1 c_2^* + c_1^* c_2) \right. 
\nonumber \\
 & & \left.  - \frac{1}{2} {\rm csch}(\gamma\pi) (1+ 2n_k)
  (c_1 c_2^* - c_1^* c_2) \right] 
\; ,
\nonumber\\
A_2 &=& -\frac{\pi}{k} e^{-\gamma \pi} \left[(1+ {\rm csch}^2
(\gamma\pi)((1+2 n_k)
|c_2|^2 + n_k)  - \frac{1}{2} {\rm csch}^2(\gamma\pi) (1+ 2n_k) 
(c_1 c_2^* + c_1^* c_2) \right]
\; ,
\nonumber\\
A_3 &=& \frac{\pi}{2 k} e^{-\gamma \pi} \left[{\rm coth}(\gamma\pi) 
{\rm csch}
(\gamma\pi)
  ((1+2 n_k)|c_2|^2 + n_k) - \frac{1}{2} {\rm coth}(\gamma\pi)
{\rm csch}(\gamma\pi) 
  (1+ 2n_k) (c_1 c_2^* + c_1^* c_2) \right. \nonumber \\
  &  & \left. + \frac{1}{2} {\rm csch}(\gamma\pi) (1+ 2n_k)
  (c_1 c_2^* - c_1^* c_2) \right] \; .  
\label{eq:A123im}
\end{eqnarray}
The last terms in the expressions for $A_1$ and $A_3$ are imaginary.  Substitution into
Eqs. (\ref{eq:I00}) and (\ref{eq:I}) shows that the contributions of these terms to the energy-momentum tensor cancel.
 
Noting that $\nu$ is imaginary, one sees that the argument for the
vanishing of the first integral in Eq. (\ref{eq:thrI}) is unchanged.  Since the
terms in the integrand are being bounded for the second integral one
must take the real part of $\beta_i$ and substitute that for $\beta_i$
in Eqs. (\ref{eq:J12}) and (\ref{eq:J12value}).  After making this substitution it is still the case that each term in (\ref{eq:J12value}) vanishes in the limit $\eta \rightarrow 0^-$.  The argument for the vanishing of the third integral in Eq.(\ref{eq:thrI}) is unchanged when $\nu$ is imaginary.  Thus, the proof is  valid for imaginary values of $\nu$.

For $\nu = n= 0,1$ we use the identity 
\begin{eqnarray}
 H_n^{(1)}(z) = J_n(z) + i N_n(z)
\; ,
 \nonumber \\
 H_n^{(2)}(z) = J_n(z) - i N_n(z) 
\; ,
 \label{eq:h12jn}
\end{eqnarray}
and then use the well known power series solutions for $N_0$ and $N_1$
to write
\begin{equation}
 N_n(z) = \frac{2}{\pi} J_n(z) \log(z) + P_n(z)
\; ,
\end{equation}
where $P_n(z)$ is a series of the form
\begin{equation}
P_n(z) = \sum_{j=0}^{\infty} b_{nj} z^{2j-n} \; .
\end{equation}
Because of the $\log(z)$ term it is useful to write
\begin{equation}
 \frac{{\rm d} N_n(z)}{{\rm d} z} = \frac{2}{\pi} 
\frac{{\rm d} J_n(z)}{{\rm d}z} \log(z) + Q_n(z)
\; ,
\end{equation}
with $Q_n(z)$ a series of the form
\begin{equation}
 Q_n(z) = \sum_{j=0}^{\infty} \epsilon_{nj} z^{2j-n-1} \;.
\end{equation} 
We can conclude that Eqs.\ (\ref{eq:I00}) and (\ref{eq:I}) remain the same but with
different expressions for $A_i$, $S_i$, and $\beta_i$.  The new
expressions for the $A_i$ are
given by 
\begin{eqnarray}
A_1 &=& -\frac{\pi}{2k}
 \left[\left(1 + \frac{4}{\pi^2} (\log z)^2 \right) ((1+2 n_k)|c_2|^2 + n_k)
     + \frac{1}{2} \left(1 - \frac{4}{\pi^2} (\log z)^2 \right)(1+2 n_k) (c_1 c_2^* + c_1^* c_2)
  \right. \nonumber \\
 &  & \left.    +  \frac{2 i}{\pi} \log z \,(1+ 2n_k) (c_1 c_2^* - c_1^* c_2)\right] 
\; ,
\nonumber \\
A_2 &=& -\frac{\pi}{k} \left[\frac{2}{\pi} \log z \, ((1+2 n_k)|c_2|^2 + n_k) 
       - \frac{1}{\pi} \log z \,(1+2 n_k) (c_1 c_2^* + c_1^* c_2) 
        + \frac{i}{2} (1+ 2n_k) (c_1 c_2^* - c_1^* c_2)\right] 
\; ,
\nonumber \\
A_3 &=& -\frac{\pi}{2k} \left[((1+2 n_k)|c_2|^2 + n_k)
     - \frac{1}{2} (1+2 n_k) (c_1 c_2^* + c_1^* c_2)\right] 
\;. 
\label{eq:A123int}
\end{eqnarray}
Note that the $A_i$ are now functions of both $k$ and $\log(z)$.  The
new expressions for the $\beta_i$ and $S_i$ are given in Table 2.
\hfil\break
\vspace{.4cm}
\begin{center}
\begin{tabular}{|c||c|c|} \hline 
\rule[-0.3cm]{0mm}{0.80cm}
 $\;\;i\; \; $ &$\; \; \; \; \beta_i \; \; \; \; $&$\; \; \;  S_i\; \; \;$
 \\ 
\hline
\hline
\rule[-0.3cm]{0mm}{0.80cm}
1 & $5 + 2n$ & $z^5 J_n^2(z)$ \\ \hline 
\rule[-0.3cm]{0mm}{0.80cm}
2
& $5 $ & $z^5 J_n(z) P_n(z)$ \\ \hline 
\rule[-0.3cm]{0mm}{0.80cm}
3 & $5 - 2n$ & $z^5 P_n^2(z)$
\\ \hline 
\rule[-0.3cm]{0mm}{0.80cm}
4 & $3 + 2n$ & $z^3 J_n^2(z)$ \\ \hline 
\rule[-0.3cm]{0mm}{0.80cm}
5 & $3 $ & $z^3
J_n(z) P_n(z)$ \\ \hline 
\rule[-0.3cm]{0mm}{0.80cm}
6 & $3 - 2n$ & $z^3 P_n^2(z)$ \\ \hline 
\rule[-0.3cm]{0mm}{0.80cm}
7 &
$3 + 2n$ & $z^4 \frac{\rm d}{{\rm d}z} J_n^2(z)$ \\ \hline 
\rule[-0.3cm]{0mm}{0.80cm}
8 & $3 $ & $z^4
\left(\frac{\rm d}{{\rm d}z} J_n(z)\right) P_n(z) + J_n(z) Q_n(z))$ \\ \hline
\rule[-0.3cm]{0mm}{0.80cm}
9 & $3 - 2n$ & $2 z^4 P_n(z) Q_n(z)$ \\ \hline 
\rule[-0.3cm]{0mm}{0.80cm}
10 & $3 + 2n$ & $z^5
\left(\frac{\rm d}{{\rm d}z} J_n(z)\right)^2 $ \\ \hline 
\rule[-0.3cm]{0mm}{0.80cm}
11 & $3 $ & $z^5
\left(\frac{\rm d}{{\rm d}z} J_n(z)\right) Q_n(z)$ \\ \hline 
\rule[-0.3cm]{0mm}{0.80cm}
12 & $3 - 2n$ &
$z^5 Q_n^2(z)$ \\ \hline
\end{tabular}
\end{center}
\begin{center}
{Table 2}
\end{center}
Substitution these expressions into Eqs. (\ref{eq:I00}) and (\ref{eq:I}) shows that for the first integral in (\ref{eq:thrI}) the expressions are of the same form, except that some terms have factors of $\log(z)$.  However these do
not prevent the terms in the first integral on the right hand side of
(\ref{eq:thrI}) from vanishing asymptotically.  For the second integral there
are still terms of the form given in Eq. (\ref{eq:J12}).  However, some of them
also have factors of $\log(z)$.  Inserting factors of $\log(z)$ into Eq. (\ref{eq:J12value}) and computing the integrals, one finds that the terms all vanish in the limit $\eta \rightarrow 0^-$.  The factors of $\log(z)$ also do not affect the asymptotic vanishing of the third integral on the right in Eq. (\ref{eq:thrI}).
     
Therefore, in all cases where $\Re (\nu) < {3\over 2}$ the quantity $\langle
T_{ab} \rangle_{SD}$ vanishes in the limit $\eta \rightarrow 0^-$.

\section{The Harmonic Oscillator with $\omega \rightarrow 0$}

In this Appendix we give a detailed discussion of the
harmonic oscillator with vanishing frequency, pointing
out the analogy with the $\nu \rightarrow {3\over 2}$ limit
in de Sitter space. Consider a simple harmonic oscillator with Hamiltonian
\begin{equation}
H = {1\over 2}\dot \phi^2 + {1\over 2} \omega^2\phi^2\,.
\end{equation}
The single degree of freedom $\phi$ can be quantized by introducing
the operator representation
\begin{equation}
\phi (t) = \langle\phi (t)\rangle + a\psi(t) + a^{\dagger}\psi^*(t)\,,
\end{equation}
where $a^{\dagger}$ and $a$ are creation and destruction operators
obeying
\begin{equation}
[a, a^{\dagger}] = 1\,.
\end{equation}
The canonical commutation relations $[\phi, \dot \phi] = i$ are
satisfied provided the mode function $\psi$ obeys the Wronskian
condition (\ref{eq:wronskian}).  The equation of motion for the
Heisenberg operator $\phi (t)$ implies that the mode functions also
satisfy
\begin{equation}
\ddot \psi + \omega^2 \psi = 0\,.
\label{eom}
\end{equation}
In order to satisfy the Wronskian condition we choose the fundamental normalized
complex solution to this equation to be
\begin{equation}
f (t)\equiv {1 \over \sqrt{2\omega}}\ e^{-i\omega t}\,.
\label{fmode}
\end{equation}
The general solution satisfying both the Wronskian condition and (\ref{eom})
can be written as the linear combination
\begin{equation}
\psi (t) = \alpha f (t)+ \beta f^*(t)\,,
\end{equation}
where
\begin{equation}
|\alpha|^2 - |\beta|^2 = 1\,.
\end{equation}
Thus, up to an irrelevant overall phase, the Bogoliubov coefficients
can be parameterized in terms of two real parameters in the form
\begin{eqnarray}
\alpha &=& \cosh\theta\,,\nonumber\\
\beta &=& \sinh\theta\ e^{-i\delta}\,.
\label{ABog}
\end{eqnarray}

We could keep the c-number expectation value $\langle\phi(t)\rangle$
non-zero in general, which would lead to the general Gaussian state with
displaced origin that we considered in some earlier papers
\cite{HKMP}. Since we do not require it in our free field de Sitter
calculation, we specialize to the case where $\langle\phi(t)\rangle =
\langle\dot\phi(t)\rangle = 0$ in the following. With this restriction
the field operator can be written simply as
\begin{equation}
\phi (t)= a\psi(t) + a^{\dagger}\psi^*(t)\,.
\end{equation}
We make use of the freedom in the $\alpha$ and $\beta$ coefficients to
require
\begin{eqnarray}
\langle aa\rangle &=& \langle a^{\dagger} a^{\dagger}\rangle = 0 
\; ,
\nonumber\\
\langle a a^{\dagger}\rangle &=& \langle a^{\dagger}a\rangle 
+\ 1 = {\sigma + 1\over 2}\,,
\end{eqnarray}
with no loss of generality. The positive real parameter $\sigma \ge 1$
has the interpretation $\sigma = 1 + 2 n$, with $n$ the average number
of ``particles" in the $a$ basis. Thus we need the three parameters $\theta$,
$\delta$, and $\sigma$ to specify the general semi-classical (coherent) state of the field. 

Using the definitions above it now follows that the average energy in this state is
\begin{equation}
E = \langle H\rangle = {\omega\over 2}\ \sigma\ (1 + 2|\beta|^2)\,.
\end{equation}
Note that we have {\it not} insisted that the state be an eigenstate of
the number operator $a^{\dagger} a$ so $n$ is only the {\it average}
number of particles in the state, which can take on any non-negative
value and need not be an integer. If $n > 0$ and the state is a Gaussian, consistent with all connected correlations vanishing except the two-point function, then it turns out that the state is necessarily a mixed state \cite{HKMP}.

We are interested in the nature of the ``vacuum" state in the limit $\omega
\rightarrow 0$, {\it i.e.} when our single degree of freedom becomes
that of a {\it free} particle. In this limit we can no longer retain
the complex oscillating mode function basis since their normalization
diverges due to the $(2\omega)^{-{1\over 2}}$ factor in
(\ref{fmode}). However, the mode equation $\ddot \psi =0$ clearly
possesses the regular {\it real} solutions $u =1$ and $v=t$, which we
can obtain from the complex oscillatory solutions by taking
appropriate linear combinations in the limit of vanishing frequency,
{\it i.e.}
\begin{eqnarray}
u(t) &\equiv & \lim_{\omega \rightarrow 0} \left\{ \sqrt{\omega\over 2}
(f + f^*)\right\} = 1\,,\nonumber\\
v(t) &\equiv & i\ \lim_{\omega \rightarrow 0} \left\{{1\over \sqrt{2\omega}}
 (f - f^*)\right\} = t\,.
\label{Auvdef}
\end{eqnarray}
These definitions are analogous to Eq.\ (\ref{uvdeS}) of the text, with $\omega$ replacing $3-2\nu$.

The general linear combination of modes can then be rewritten in the form
\begin{equation}
\psi (t) = {(\alpha + \beta) \over \sqrt{2\omega}} u
- i {\sqrt \omega \over 2}(\alpha - \beta) v = Bu + Av \rightarrow  At + B\,,
\end{equation}
where the quantities
\begin{eqnarray}
A &\equiv& -i\ \lim_{\omega \rightarrow 0} \left\{ \sqrt{\omega\over 2} (\alpha - \beta)
\right\}
\; ,
\nonumber\\
B &\equiv& \lim_{\omega \rightarrow 0} \left\{ {1\over\sqrt{ 2\omega}} (\alpha + \beta)
\right\}
\; ,
\label{ABcoef}
\end{eqnarray}
are analogous to those defined by (\ref{ABdeS}) of the text. They also satisfy
\begin{equation}
A^*B - B^*A = i (|\alpha|^2 - |\beta|^2) = i\,.
\label{ABcond}
\end{equation}
We can also define the time-independent Hermitian operators
\begin{eqnarray}
Q &\equiv &  {1\over\sqrt{ 2\omega}}\left[(\alpha + \beta)\ a +
(\alpha + \beta)^*\ a^{\dagger}\right] \rightarrow Ba + B^* a^{\dagger}\,,\nonumber\\
P &\equiv & -i \sqrt{\omega\over 2}\ \left[(\alpha - \beta)\ a
- (\alpha - \beta)^*\ a^{\dagger}\right] \rightarrow Aa + A^* a^{\dagger}\,,
\end{eqnarray}
analogous to (\ref{eq:QPdef}) in the de Sitter case. They
 obey the canonical commutation relations
\begin{equation}
[Q, P] = i\,,
\label{com}
\end{equation}
so that in the limit $\omega \rightarrow 0$
\begin{equation}
\phi (t) \rightarrow u(t) Q + v(t)P = Q + tP\,,
\end{equation}
and in the same limit
\begin{equation}
E = {1\over 2} \langle P^2\rangle = {\sigma \over 4}\lim_{\omega \rightarrow 0}
\left\{\omega \ |\alpha - \beta|^2\right\} = {\sigma \over 2} |A|^2\,.
\label{ener}
\end{equation}
Since $\sigma$ is just an overall factor and we are interested in pure
vacuum-like states, we can set $\sigma = 1$.

We see that depending on exactly how we take the limit $\omega
\rightarrow 0$, we can get many different values for the energy. If we
tried to keep the pure positive frequency solution $\psi = f$, {\it
i.e.} $\alpha =1$ and $\beta = 0$, then the energy would vanish in
the limit $\omega \rightarrow 0$. However, by computing the
correlation function, $\langle \phi (t)\phi(t')\rangle$, we would find
that the correlator {\it diverges} in this limit.  In our simple
example it is clear why. The ordinary ground state vacuum wave
function of the simple harmonic oscillator is a Gaussian with a width
proportional to $\omega^{-{1\over 2}}$. Hence $\langle Q^2\rangle \sim
\omega^{-1}$ in this state, which becomes non-normalizable in the
limit, unless we were to put the particle in a box of finite
volume. But this option is unavailable to us if the ``particle'' is a
mode of a field and the range of the field variable $\phi$ is
$(-\infty, \infty)$. Thus, we must reject the positive frequency
harmonic oscillator vacuum in the limit $\omega \rightarrow 0$ since
it is a non-normalizable state.

In order for the state to remain normalizable and the mode functions
bounded, the quantities $A$ and $B$ must remain finite in the limit of
vanishing $\omega$.  Inspection of the definitions (\ref{ABcoef})
shows that this requires that the original Bogoliubov coefficients
$|\alpha|$ and $|\beta|$ go to infinity. In fact, using (\ref{ABog})
in the definitions of $A$ and $B$ a simple calculation shows that the
Bogoliubov parameters $\theta$ and $\delta$ must behave like
\begin{eqnarray}
\theta &\rightarrow &-{1\over 2}\log \left({\omega\over 2}\right) +
\log \vert A\vert\,,\nonumber\\
\delta &\rightarrow &\pi + \omega T
\; ,
\end{eqnarray}
as $\omega \rightarrow 0$, so that
\begin{eqnarray}
A &=& -i\vert A \vert\,,\qquad {\rm and}\nonumber\\
B &=& {1\over 2 \vert A \vert} + {i\over 2} \vert A\vert\ T
\; ,
\end{eqnarray}
remain finite in that limit. Thus, the two complex numbers $A$ and
$B$ obeying (\ref{ABcond}) and the state they characterize can be
specified by the two real numbers $|A| \ge 0$ and $T$, up to an
unobservable overall phase.  Comparison of these expressions with the
corresponding equations (4.7) and (4.13) in the paper by Allen and
Folacci~\cite{a-f} shows that our $A$ here is equivalent to Allen and Folacci's $2A$.
This is because of a factor of 2 difference in the normalization of their 
time dependent mode $v(t)$.

Since $\theta \rightarrow \infty$, in a sense the normalizable
``vacuum'' state with finite $\langle P^2\rangle = |A|^2$ and finite
$\langle Q^2\rangle = |B|^2$ is infinitely far from the usual ground
state vacuum of the harmonic oscillator, and is {\it not} an
eigenstate of the free particle Hamiltonian at $\omega = 0$. Instead,
it is just a Gaussian state centered at $\langle Q\rangle = 0$ and
$\langle P \rangle = 0$ with normalized wave function \cite{HKMP}
\begin{equation}
\Psi (q) = {1 \over (2\pi |B|^2)^{1\over 4}}\exp 
\left( - {|A|\over 2 B^*}\ q^2 \right)
\; ,
\label{gaus}
\end{equation}
in the position representation where $Q\Psi = q\Psi $ and $P\Psi = -i
{{\rm d}\over {\rm d}q}\Psi$.  This state is time reversal invariant
if and only if $T = 0$, in which case $B = {1 \over 2 |A|}$ becomes
real. On the other hand, an eigenstate of the free Hamiltonian $P^2/2$
is necessarily non-normalizable since diagonalizing $P$ requires
infinite spread in $Q$. These remarks carry over equally well to
the $k=1$ mode in the de Sitter case.

There is no reason to diagonalize $H$ in the field theory case as
Allen and Folacci do in their Eqs. (4.26) to (4.28)~\cite{a-f}, since
the energy-momentum tensor is time dependent in general, and
back-reaction has been neglected in our simple calculations.
Diagonalizing just the matter Hamiltonian without taking into account
the metric is no better than a leading order mean field approximation
to the much more difficult full problem with metric fluctuations taken
into account. Given that what we are doing is only a semi-classical
approximation to this full problem in any case, what makes the most
sense from our perspective is to take well-defined normalizable
initial adiabatic states for the scalar field and allow them to evolve
as they will. Our initial state trial wave functional (\ref{gaus})
then remains a bounded Gaussian functional, consistent with
semi-classical mean field methods (although it may well spread over
time depending on the dynamics).

As should be clear, the non-normalizable positive frequency vacuum
with $\alpha =1$ and $\beta =0$ is analogous to the Bunch-Davies vacuum,
while the normalizable state related to it by an infinite Bogoliubov
transformation (but only a finite shift in zero point energy) is
analogous to the Allen-Folacci states parameterized by $A$ and $B$ or
$|A|$ and $T$. However, whereas the energy of the normalizable
state is finite and arbitrary, depending on the arbitrary $|A|^2$
coefficient in (\ref{ener}), and the non-normalizable positive
frequency state has zero energy in this simple harmonic oscillator
example, the converse is true at late times in de
Sitter space. This is due to the different kinematics and form of the
energy-momentum tensor of the scalar field in the de Sitter case.
Comparison of the energy (\ref{ener}) of the harmonic oscillator
example and the corresponding $k=1$ late time energy density in the de Sitter case 
with $m=0$ and $\xi$ small, {\it viz.} 
\begin{equation}
\varepsilon_1 \rightarrow {1\over 2a^2} \left({{\rm d}\phi_1\over {\rm d}\eta}
\right)^2
+ 3 \xi \phi_1^2 =  {1\over 4\pi^2 a^2}\left({{\rm d}v\over {\rm d}\eta}
\right)^2
P^2 + {3 \xi\over 2 \pi^2} (Q + vP)^2
\; ,
\label{eq:epsosc}
\end{equation}
shows that the difference is the redshift factor of $a^{-2}$
multiplying the infrared finite $\langle P^2\rangle$ in the de Sitter
case. Thus, whereas this can take on the finite state dependent value
$|A|^2$ in the harmonic oscillator example, its asymptotic late time value
vanishes in de Sitter space when multiplied by $a^{-2}$, and becomes state 
independent in any infrared finite state. Further, in the simple harmonic 
oscillator the usual ground state becomes an eigenstate of $P$ and has zero 
energy as $\omega \rightarrow 0$, since $\omega^2 \langle (Q+vP)^2 \rangle \sim
\omega \rightarrow 0$ drops out of the energy in this limit, whereas
in the de Sitter case, since $\xi$ goes to zero only {\it linearly}
with $3 - 2 \nu$, the Bunch-Davies expectation value receives a finite
contribution from the $k=1$ mode from the infinite spread in $\xi
\langle Q^2 \rangle$ which has no $a^{-2}$ redshift factor in (\ref{eq:epsosc}).

\end{document}